\documentclass[fleqn,usenatbib]{mnras}

\usepackage{newtxtext}

\usepackage[T1]{fontenc}

\DeclareRobustCommand{\VAN}[3]{#2}
\let\VANthebibliography\thebibliography
\def\thebibliography{\DeclareRobustCommand{\VAN}[3]{##3}\VANthebibliography}

\usepackage{graphicx}	
\usepackage{amsmath}	
\usepackage{amssymb}	

\usepackage{amsfonts}
\usepackage{url}

\newcommand{\tx}{\textit}
\newcommand{\beq}{\begin{equation}}
\newcommand{\eeq}{\end{equation}}
\newcommand{\bdi}{\begin{displaymath}}
\newcommand{\edi}{\end{displaymath}}
\newcommand{\beqn}{\begin{eqnarray}}
\newcommand{\eeqn}{\end{eqnarray}}

\newcommand{\der}{{\rm d}}
\renewcommand\({\left(}
\renewcommand\){\right)}

\newcommand{\Lp}{\emph{Left panel}}
\newcommand{\lp}{\emph{left panel}}
\newcommand{\rp}{\emph{right panel}}
\newcommand{\Rp}{\emph{Right panel}}
\newcommand{\Mpuls}{\dot{M}_{\rm pulse}}
\newcommand{\vesc}{v_{\rm esc}}
\newcommand{\fq}{f_Q (\rhos)}

\newcommand{\gcmc}{\rm{g/cm}^3}

\newcommand{\MS}{M_{\odot}}
\newcommand{\RS}{R_{\odot}}
\newcommand{\rhv}{\rho_{\rm vir}}
\newcommand{\rv}{r_{\rm vir}}
\newcommand{\Mv}{M_{\rm vir}}
\newcommand{\rhos}{\rho_\star}
\newcommand{\rhot}{\rho_{\rm cut}}
\newcommand{\Ms}{M_\star}
\newcommand{\Teff}{T_{\rm eff}}
\newcommand{\rhodm}{\rho_\chi}
\newcommand{\mdm}{m_\chi}

\newcommand{\Qdm}{\hat{Q}_{\rm DM}}

\newcommand{\parrdr}[1]{\partial_r \, (\der_r #1)}
\newcommand{\parddr}[1]{\partial_{\rhos} \, (\der_r #1)}

\title[First dark stars]{Stability and pulsation of the first dark stars}

\author[Rindler-Daller et al.]{
Tanja Rindler-Daller,$^{1}$\thanks{E-mail: tanja.rindler-daller@univie.ac.at}
Katherine Freese,$^{2,3}$
Richard H.D. Townsend,$^{4}$
Luca Visinelli$^{5}$
\\
$^{1}$Institut f\"ur Astrophysik, Universit\"atssternwarte Wien, University of Vienna, 1180 Vienna, Austria \\
$^{2}$Department of Physics, The University of Texas at Austin, Austin, 78712 TX, USA\\
$^{3}$Oskar Klein Center for Cosmoparticle Physics, University of Stockholm, 10691 Stockholm, Sweden\\
$^{4}$ Department of Astronomy, University of Wisconsin, Madison, 53706 WI, USA\\
$^{5}$ GRAPPA, Institute for Theoretical Physics and Delta Institute, University of Amsterdam, 1098 XH Amsterdam, The Netherlands
}

\date{Accepted XXX. Received YYY; in original form ZZZ}

\pubyear{2015}

\begin{document}
\label{firstpage}
\pagerange{\pageref{firstpage}--\pageref{lastpage}}
\maketitle

\begin{abstract}
The first bright objects to form in the Universe might not have been ``ordinary'' fusion-powered stars, but ``Dark Stars'' (DSs) powered by the annihilation of dark matter (DM) in the form of Weakly Interacting Massive Particles (WIMPs). If discovered, DSs can provide a unique laboratory to test DM models. DSs are born with a mass of order $M_\odot$ and may grow to a few million solar masses; in this work we investigate the properties of early DSs with masses up to $\sim \! 1000 \, M_\odot$, fueled by WIMPS weighing $100$ GeV. 
We improve the previous implementation of the DM energy source into the stellar evolution code MESA. We show that the growth of DSs is not limited by astrophysical effects: DSs up to $\sim \!1000 \, \MS$ exhibit no dynamical instabilities; DSs are not subject to mass loss driven by super-Eddington winds. We test the assumption of previous work that the injected energy per WIMP annihilation is constant throughout the star; relaxing this assumption does not change the properties of the DSs. Furthermore, we study DS pulsations, for the first time investigating non-adiabatic pulsation modes, using the linear pulsation code GYRE. We find that acoustic modes in DSs of masses smaller than $\sim \! 200 \, M_\odot$ are excited by the $\kappa-\gamma$ and $\gamma$ mechanism in layers where hydrogen or helium is (partially) ionized. Moreover, we show that the mass loss rates potentially induced by pulsations are negligible compared to the accretion rates. 
\end{abstract}

\begin{keywords}
dark matter; astroparticle physics; stars: evolution; stars: oscillations (including pulsations); dark ages, reionization, first stars
\end{keywords}



%
\section{Introduction}\label{sec:Intro}
%

The formation of the first stars in the Universe marks a turning point in cosmic history,
with implications for reionization, metal enrichment of the intergalactic medium, and formation of the first galaxies. In order to understand these processes, it is crucial to investigate the stellar physics and nature of the first generation of stars. In the standard cosmological model with a cosmological constant $\Lambda$ and cold dark matter (DM), the formation of the first stars is believed to take place within minihalos of masses $\sim 10^{6} - 10^{8} M_{\sun}$ at redshifts of $z \sim 10-50$. The gravitational potential of these minihalos is dominated by DM, contributing $\sim 85 \%$ of the total mass.  Baryons -- providing the other $\sim 15 \%$ of the minihalo mass -- are gravitationally pulled into the centers of the minihalos, where they start to dominate the central potential well and this central region heats up. As a result, baryons radiate away their energy, cool off and are able to become cool enough to form the first stars in those centers. However, the details of this overall picture are complicated, and the subject has branched into so many different aspects over the years, that we cannot give a proper account of all the relevant literature here. Instead,
we will focus on a particular aspect, namely the impact of DM particle physics onto the stellar evolution of so-called ``dark stars''. In fact, this paper connects onto previous works which have focused on the question of how DM impacts the first stars.

As the first stars form at high redshifts, in the gravitationally contracted centers of DM minihalos, the high ambient DM densities, $\rhodm$, provide the potential for DM particles to significantly alter the chemistry and physics of the formation and evolution of the first stars at these redshifts.
Among the canonical cold dark matter candidates are weakly-interacting massive particles (WIMPs) which, in many particle theories, are their own antiparticles and can therefore annihilate with each other. This annihilation process in the early Universe is able to leave the correct DM relic abundance today, commonly referred to as the ``WIMP miracle''. But annihilation becomes also particularly important in regions with high DM density, $\rhodm$, as the annihilation rate of the DM particles scales with the density squared (i.e. $\Gamma_{\rm{ann}} \propto \rhodm^2$). These high density regions are exactly those sites where the first stars are forming: at high redshifts ($\rhodm \sim (1+z)^3$), in the dense centers of DM minihalos. 

The work of \citealp*{Spolyar:2007qv} showed that WIMP annihilation can drastically alter the formation and evolution of the first stars. Above a certain baryonic density threshold, DM annihilation products remain trapped in the star, thermalize and provide a heat source that dominates over all cooling mechanisms. Subsequently, the protostellar cloud will continue to contract at a slower rate until hydrostatic equilibrium is reached and a ``dark star'' (\textbf{DS}) is born. In this work, we focus on the study of dark stars (\textbf{DSs}) fueled by WIMPs with mass $\mdm = 100$ GeV. Earlier studies of the effect of different WIMP masses on the properties of DSs can be found in \citealp*{Freese08}, \citealp*{Freese10} for masses ranging between $\mdm = 1$ GeV and $\mdm = 10$ TeV.  

It is important to stress that the first DSs are made primarily of hydrogen and helium and less than $\sim \! 0.1 \%$ of the mass is contributed by DM. Nevertheless, they shine solely due to DM heating. The term ``dark'' refers to the power source, and not the brightness or the primary matter constituent.  A detailed review on DSs can be found in \citealp*{Freese:2015mta}.

DSs are born with masses $\sim (1-10) \, M_\odot$ and then can grow 
by accreting material from the surrounding halo, as studied by  \citealp*{Freese08}, \citealp*{Freese10}. These studies determined that DSs are bright ($> 10^6 L_{\sun}$), of low density ($\rhos \lesssim 10^{-6}$ g/cm$^3$), large ($\gtrsim \! 10$ AU) and cool ($T_{\rm{eff}} < 10,000$ K). Thanks to the low surface temperatures, early DSs do not suffer feedback mechanisms which could prevent the accretion of further material onto the star (such feedback processes were already studied in \citealp*{TM}), a limitation which fusion-powered stars usually face. That property is a key to enable DSs to keep growing until there is no further DM fuel. Therefore, they can grow to supermassive size. If DSs grow to $>\mathcal{O}(10^5 M_{\sun})$, they reach luminosities of order $(10^9-10^{11}) L_{\sun}$ (\citealp*{Freese10,TRD15}, henceforth abbreviated \textbf{F10} and \textbf{RD15}, respectively). These high luminosities puts them into the observable limits of next-generation telescopes, such as the James Webb Space Telescope (JWST) \citealp*{Gardner:2006ky}, as was shown in F10 and \citealp*{Ilie12}. 
In order to sustain the energy support via DM annihilation over an extended period of time, there are two mechanisms that can enhance the DM density in the center of the DS: (i) gravitational contraction (\citealp*{Spolyar:2007qv,Freese09}), and (ii) WIMP capture (\citealp*{Iocco08,Freese08b}).  These mechanisms can potentially supply DM resources  to provide the necessary fuel for the growth of supermassive DSs.

In the first mechanism, DM is pulled into the center of the halo, where the DSs form, by the gravitational attraction of baryons. This generic process is expected to occur in the halos in which the first DSs form\footnote{DM is responsible for the formation of halos, pulling the baryons into the halo centers. However, later when baryons dominate those centers, they pull in more DM into the central regions in turn, again due to gravity ("gravitational contraction").}. To a good approximation, the DM particles can be assumed to follow the infalling baryons adiabatically (commonly treated via adiabatic contraction (AC), see \citealp*{Blumenthal})\footnote{The goodness of the AC approximation for the formation of DSs was tested in \citealp*{Freese09}. They found that the difference in the contracted DM density profile compared to an exact treatment, using gas profiles from the first-star simulations of \citealp*{TM}, stay within a factor of two.}. The AC boosts the central DM densities by many orders of magnitude, resulting in DM densities high enough to power a DS\footnote{While there are still uncertainties about the exact inner density profile of a DM halo, the formation of a DS was shown to occur regardless of the detailed assumptions on the initial profiles and orbits (\citealp*{Freese09,NTO09}). Even in the extreme case of a cored Burkert profile, DSs were still found to form.}. %
Studies of the dynamics of cold, collisionless DM within galactic halos have shown that the central DM abundance is replenished due to a continuous infall of DM on centrophilic orbits, 
see e.g. \citealp*{GS81, Gerhard1985, VM98, Valluri10, Valluri12}. Given the self-similarity of DM dynamics on all galactic scales, one can assume that the same replenishment applies within the minihalos where DSs form and reside. As in previous works, \citealp*{Freese08}, F10, RD15, we implement AC using the Blumenthal method. However, to increase numerical stability in the calculations, in this paper, we improve the implementation of RD15 by replacing numerical with analytical derivatives.  

At a later stage in the evolution of a DS, once there is no longer sufficient DM fuel due solely to gravitational effects, the DS contracts and becomes dense enough that a second effect may become important: the capture of WIMP particles may serve to replenish the DM reservoir inside the star.  DM particles from the surrounding environment elastically scatter off of nuclei inside the star and get trapped (\citealp*{Freese08b, Iocco08}).   
While the existence of DSs relies on WIMP annihilation, WIMP capture relies upon a different process, namely the elastic scattering of WIMPs on nucleons. 
The details of the process and the value of the scattering cross section are currently unknown and are actively searched for in direct WIMP detection experiments, see e.g. \citealp*{D0,D1,D2,D3,D4,D5}. The importance of this capture process for fueling DSs depends upon the scattering cross section and ambient WIMP density. If WIMP capture is included into the analysis, DSs tend to be denser and hotter compared to the cases in which they formed via AC alone, see e.g. F10. In the present paper, we have decided not to consider capture, which e.g. could be due to a small WIMP-nucleon cross section several orders of magnitude lower than current experimental bounds. 

We wish to briefly comment on work of other authors on the subject of DS formation.  For example, \citealp*{smith} followed the collapse of the protostellar cloud to hydrogen density $10^{14}$/cm$^3$, not high enough for the required  density of $10^{17}$/cm$^3$ for a DS to form in hydrostatic and thermal equilibrium in the case of 100 GeV WIMPs.  Further collapse beyond the values reached by \citealp*{smith} is required for onset of DSs (and their work does not imply that DSs are unable to form).
In fact, it is argued in \citealp*{arxiv} that the work of \citealp*{smith} supports the existence of DSs (although the simulation is not refined enough to follow the inner parts where the DS would form), because heating due to DM annihilation would actually stabilize the protostellar cloud against fragmentation.
Now, if the assumption of a static, spherically-symmetric DM profile is relaxed and a self-consistent DM distribution is adopted, gas collapse is not halted by DM annihilation until the protostar is formed. In \citealp*{stacy}, the collapse is followed down to scales of about 5 AU, without studying whether the collapsed object is a DS or an ordinary protostar. It is argued there that the DM reservoir available for annihilation inside a DS lasts for a lifetime of the DS which is only a few thousand years. However, as mentioned above, DM particles follow centrophilic orbits, continuously falling into the central region. This enables a large fraction of DM to come into the DS from far outside the small region near the halo center where the DS resides. The above simulations only considered this innermost region. Hence, we disagree with the claims made in \citealp*{stacy} that DSs should not be long-lived.
Recent review papers on first star formation, such as e.g. \citealp*{Klessen, Haemmerle}, include some discussion on DSs, cautioning that the formation process remains unsettled.  Surely, studies of DS formation require more simulations in the future to accurately model the details of the formation process. Our work follows the stellar evolution of DSs, once they reach equilibrium.

Early works on the stellar evolution of DSs were based on the assumption of polytropic stellar interiors (\citealp*{Freese08}; F10). In order to study the evolution and structure of DSs beyond this assumption, the 1D stellar evolution code MESA\footnote{http://mesa.sourceforge.net/} (``Modules for Experiments in Stellar Astrophysics'') (see \citealp*{MESAI,MESAII,MESAIII}) was used in RD15. In MESA the stellar structure equations are solved self-consistently such that no restrictive assumptions on the structure of the stellar models or the equations of state have to be made. To extend the capabilities of MESA to DSs, we implemented in RD15 an additional module that locally derives and adds the energy from DM annihilation self-consistently.

The comparison of basic stellar properties between the polytropic and MESA models in RD15\footnote{Without considering DM capture in both cases.} showed that the results agree qualitatively very well (deviations of basic parameters being within factors of a few). All of these studies on the properties of DSs were based on the assumption that the fraction of energy per WIMP annihilation deposited into the DS is independent of the density and constant throughout the whole star.  

However, this assumption does not hold in general: as the star grows larger, the stellar densities in the outer regions decrease. This also means that the cross section of WIMP annihilation products with the baryonic stellar medium decreases. Annihilation products that are created near the surface might therefore leave the star without any interaction, or after depositing only a small fraction of their energy into the star.  In this paper, we compare stellar models with different treatments of the injected energy rate into the low-density surface regions. Our tests show that the energy injected into these regions is indeed negligible compared to the energy produced in the dense core.  

Besides this test of fundamental assumptions, we also extend the study of basic properties of DSs in their early evolutionary stages, as follows. 

We show that the growth of DSs is not halted by stellar physical processes by analyzing the stability of our models. Here, we focus on dynamical instabilities that can potentially disrupt the equilibrium structure of a star and lead to a core collapse.

Another important effect, that could strongly alter the evolution of a star, is the so-called Eddington limit. If a star
becomes so bright that this limit is exceeded, the radiation pressure of the outflowing photons dominates over the gravitational inwards-pull. This can potentially lead to mass outflows, i.e. super-Eddington winds. Since these can lead to mass loss in ordinary, 
massive stars, see e.g. \citealp*{langer1997eddington,Graefener11,Sanyal15,Quataert15}, we explicitly test whether that is also the case in DSs. In fact, we will show that neither effect prevents DSs from growing further in mass.

Moreover, we investigate non-adiabatic stellar oscillations. In an adiabatic calculation, RD15 found that acoustic -- or p-modes -- can be present in DSs. The pulsation periods were shown to be in a range between a few and several thousand days in the star's rest frame -- depending upon parameters like WIMP and stellar mass. However, that adiabatic analysis was not sufficient to determine which modes would be actually excited. We tackle this question here by using the non-adiabatic module of the pulsation code GYRE, \citealp*{Gyre}, which is a numerical code that solves for both the adiabatic and non-adiabatic modes of stellar pulsations. Indeed, our results suggest that, below a certain mass limit ($\sim \! 200~ \MS$), the $\kappa-\gamma$ and $\gamma$ mechanisms can excite pulsations in DSs. Moreover, we show that the upper bound on the mass loss rate induced by these pulsations stays well below the mass accretion rate. These results further support the picture that DSs can evolve to become supermassive and are not limited by instabilities or mass loss as they grow. For the case of typical accretion rates of $10^{-3} \MS/\rm{yr}$ onto the DS, our analysis indicates that such pulsations are not excited in DSs of supermassive size. 

We want to stress again that our paper does not address the formation of DSs from their primordial clouds. Instead, we assume that the DM annihilation in the surface regions of the star does not affect the formation and evolution of DSs.


This paper is structured as follows: \autoref{sec:NumMod} presents some basic equations for the evolution of DSs and their numerical modeling. This section also includes a description of our test of the energy injection assumption. In \autoref{sec:Comp}, we confirm the numerical stability of our DS models by comparing the results obtained from different codes and with different implementation details. The dynamical stability and the Eddington limit are addressed in \autoref{sec:Stab_MassLoss}. We focus on non-adiabatic pulsations in \autoref{sec:Puls}, where we identify the driving mechanism and the pulsation periods. Furthermore, we calculate an upper bound on the mass loss rate driven by the resulting pulsations. We conclude with a summary and discussion in \autoref{sec:Concl}. Appendix \ref{sec:Appendix} includes the derivation of the newly implemented analytical derivatives.

%
\section{Equations and Implementation} \label{sec:NumMod}
%


\subsection{Energy Source in Dark Stars} \label{sec:DSEq}
The energy production rate due to WIMP annihilation \tx{per unit volume} is given by
\begin{equation}
	\Qdm = \langle \sigma v \rangle m_\chi n_\chi^2 = \langle \sigma v \rangle \frac{\rho_\chi^2}{m_\chi} \, ,
	\label{eq:DMheat}
\end{equation}
where $n_\chi$ is the WIMP number density, $\mdm$ the WIMP mass, $\rhodm$ is the WIMP mass density and $\langle \sigma v \rangle$ the annihilation cross section. For the latter we use the standard value from DM freeze-out, $\langle \sigma v \rangle = 3 \times 10^{-26} {\rm cm}^3/{\rm s}$. One can see that the energy production scales as 
$\langle \sigma v \rangle / {m_\chi} $. Hence, studying a variety of WIMP masses can be traded off against studying a variety of cross sections.  In previous papers, we showed that DSs arise regardless of WIMP mass over many orders of magnitude and therefore for a wide range of annihilation cross sections.  In this paper, we restrict our work to 100 GeV WIMPs and the canonical cross section,
but our results could easily be generalized.

In MESA we need the production rate \tx{per unit mass}, which is obtained by dividing Eq.~(\ref{eq:DMheat}) by the baryonic density of the stellar gas, $\rhos$.

The luminosity arising from WIMP annihilation can be calculated with the knowledge of the annihilation end products. These are typically electrons, photons, and neutrinos. While neutrinos escape the star, the other annihilation products can be trapped inside the DS, if the conditions for DS formation are met, see \citealp*{Spolyar:2007qv}. Trapped products thermalize and heat up the star. The resulting luminosity can be calculated with
\begin{equation}
	L \sim \int \fq \, \Qdm \, {\rm d} V \, ,
	\label{eq:DMheatingLum}
\end{equation}
where ${\rm d}V$ is the volume element and $\fq$ is the fraction of the annihilation energy deposited into the star. In previous DS studies 
(e.g.~ \citealp*{Freese08}, F10 and RD15), $\fq$ was taken to be a constant, $f_Q=2/3$, throughout the whole star. This factor physically reflects the rough estimate that 1/3 of the annihilation energy is converted into neutrinos, which escape the star without any further interactions, while the remaining fraction of the energy is deposited into the star by further interactions in the stellar gas. Of course, the exact fraction deposited depends upon the particular WIMP model.

Now, the assumption that the deposited energy fraction per WIMP annihilation is constant within the whole star, $\fq = \rm{const.}$, does not hold in general: as the star grows larger, the stellar densities in the outer regions decrease. This also means that the cross section of WIMP annihilation products with the baryonic stellar medium decreases. Annihilation products that are created near the surface might therefore leave the star without any interaction, or after depositing only a small fraction of their energy into the star. This effect could lead to a feedback onto the proto-stellar cloud (during the formation), or the accretion disk (during the evolution). For instance, if a considerable amount of ionizing photons leave the star, they can cause a fragmentation of the accretion disk, preventing the star from growing and accreting material (\citealp*{TM,McKeeTan08,Stacy10,Hosokawa11b}). A detailed study of these effects during the formation as well as the evolution would require a fully self-consistent hydrodynamical simulation of not only the star, but also its environment. In addition, calculating the amount of ionizing photons from this process would require a careful study, using e.g. Pythia which follows the annihilation cascades for detailed WIMP particle models. These investigations go beyond the scope of the present paper, so we assume that possible feedback from this effect is small, i.e. we will assume that those annihilation products have no impact onto the formation of the star, nor on the further accretion of material onto the growing star. Nevertheless, we test the consequences for the stellar properties of the assumption that the energy injection rate per WIMP annihilation is independent of the stellar density. To do so, we compare stellar models with different treatments of the injected energy rate into the low-density regions. 


In reality, $\fq$ is a function of the surrounding density and of the distance to the stellar surface. In the outer regions of the star, the baryonic densities might not be high enough to absorb all the energy of the annihilation products, as the interaction cross section drops off with the gas density. Furthermore, the distance of two annihilating WIMPs to the stellar surface also affects the fraction of energy that is deposited into the star. Again, we stress that in an exact treatment, it would be necessary to focus on a particular WIMP particle (e.g. a neutralino) and trace the cascade of annihilation products in the environment where the annihilation takes place. On the other hand, most of the WIMP annihilations happen close to the center of the DSs due to the high local DM densities. The baryonic densities and the distance to the stellar surface in these regions are high enough to fully absorb the energy from all annihilation products, excluding neutrinos as shown in
\citealp*{Spolyar:2007qv}. Therefore, a deviation from $\fq = 2/3$ (for the case of 1/3 energy loss to neutrinos) is only expected in the outer envelope regions with their low densities.

For our purpose, it will be sufficient 
to get an intuition for the magnitude of the effect of $\fq$ dropping below 2/3 in the outer regions of the DS, by comparing the canonical case, $f_Q = 2/3$, with a treatment in which we set
\beq \label{eq:Heatcut}
	\fq =
	\begin{cases}
		2/3 & \text{for } \rhos > \rhot \\
		0 & \text{else.} \\
	\end{cases}
\eeq
For the values of $\rhot$ we will consider two cases, $\rhot = 10^{-8} \gcmc$ and $\rhot = 10^{-9} \gcmc$. 
Especially, the latter value reflects an estimate of the density below which the assumption of constant $\fq$ is expected to fail (\citealp*{Spolyar:2007qv}). Notice that in this treatment, it is the DM heating that is cut off below $\rhot$, but not the gas or DM density of the star.
The impact of the above choices on DSs of different masses will be discussed in \autoref{sec:Comp}. 
While an exact treatment of the induced cascades of annihilation products within the star is subject to future studies, we stress that the above approximation suffices to determine the order of magnitude of the expected impact. Indeed, our results show that the precise form of $\fq$ in the outer, low-density regions is not important for the properties of the star; the impact of applying the more conservative threshold on luminosity, radius, effective temperature and other stellar properties is less than $1\%$ for all considered masses. The details of this comparison are given in \autoref{sec:Comp}.

\subsection{Numerical Method} \label{sec:Method}
In order to solve the set of differential equations governing stellar evolution, we use the stellar evolution code MESA, release 12778 (\citealp*{NewMESA}). As in RD15, this is accomplished by implementing the additional energy source from DM heating into MESA. The \verb+run_star_extras+ module allows to include the extra terms easily through the \verb+other_energy_implicit+ option. During each time step, this extra energy is added self-consistently to the model in the same way that energy due to nuclear reactions would be. The extra routine for DSs also calculates the adiabatically contracted DM profile -- according to the Blumenthal method -- by using the baryonic density profile of the star. Once $\rhodm (r)$ is known, the DM heating rate can be calculated (cf.~Eq.~(\ref{eq:DMheat})). These routines have been adapted from \citealp*{Freese08}, F10 and RD15. However, to make sure that we obtain reliable and converged numerical results, we improve upon the previous implementation: for a fully self-consistent calculation of an extra energy source in MESA, the partial derivatives of the added heating rate with respect to density, radius and temperature have to be known. We replaced the previous difference expressions with the actual analytical formulae for the derivatives; the details of this calculation are given in Appendix \ref{sec:Appendix}.

In order to compare our findings to the relevant results in F10 and RD15, we make the same assumption here, namely that DM annihilation does not lead to a depletion of DM, i.e. we do not remove annihilated DM from the reservoir (see \autoref{sec:Intro} for the discussion of the viability of this assumption). Likewise, we choose the same accretion model, i.e. ``cold accretion''. This assumes that the entropy of the material which is accreted is equal to the entropy of the surface layers of the star, i.e. accretion does not directly heat the surface. Physically, this means that the infalling material gradually radiates away its gravitational energy when passing through the accretion disk.%
\footnote{An alternative would be to consider ``spherical accretion''. Here, shock fronts evolve at the stellar surface, increasing the entropy of the infalling material and therefore heating up the surface layers. However, \citealp*{Hosokawa10,Hosokawa12} concluded that both accretion modes lead to similar results for massive protostars above $\gtrsim 40 \,\MS$. DSs with similar mass are more extended, hence, the energy release from infalling material is even smaller. Therefore, we assume that a similar conclusion holds for DSs.}

\subsection{Initial Conditions} \label{sec:IC}
The initial conditions for the evolution of a DS are obtained by creating a so-called ``pre-main-sequence model'' in MESA. For a fixed central temperature, $T_c$, and given initial composition (zero metallicity, i.e.~$Z =0$, in DSs), the total mass depends only on the central baryonic density, $\rho_c$. By assuming a stellar structure described by an ($n=3/2$)-polytrope, a first initial guess for $\rho_c$ is obtained. This assumption is a good approximation for a fully convective star. However, the assumption of convection is relaxed during the search for a converged ``pre-main-sequence model'': the stellar structure equations, the equation of state, and the convection parameters of the mixing-length-theory (MLT) are solved until a value for $\rho_c$ is found that yields the desired initial mass. Although this procedure and the initial guess are not optimized for DSs, the models converge quickly towards an equilibrium sequence. In the present calculations, we choose an initial stellar mass of $5 \, \MS$ for the creation of the ``pre-main-sequence model''. However, we stress that this choice does not affect the conclusions in any way; choosing an initial mass of e.g. $1 \, \MS$ or $10 \, \MS$ yields identical results, once the model has converged.

In this work, we focus on some important, astrophysical aspects of the early stellar evolution of the first DSs, up to masses of about $1000 \, M_{\odot}$, although some of our results (e.g. pulsations) have been studied up to $10^4 ~\MS$. We restrict our attention in this paper to one halo environment and one WIMP particle mass of 100 GeV. The dependence of DS evolution on these latter parameters have been studied in F10 and RD15. We consider models of DSs which are accreting matter at a constant rate, $\dot M = \rm{const}$, and which formed in a host minihalo with mass $M_{\rm halo}$ at formation redshift $z_{\rm form}$ with the following choices
\begin{eqnarray}
	\dot M &=& 10^{-3} M_{\odot}{\rm yr}^{-1} \, ,  \nonumber \\ 
	M_{\rm halo} &=& 10^{6} M_{\odot} \, , \nonumber \\
	z_{\rm form} &=& 20 \, . 
\end{eqnarray}
This environment was termed ``SMH100'' in RD15 and also reflects the choice of previous studies in \citealp*{Freese08} and F10. The halo consists of $85\%$ DM and $15 \%$ baryons with a primordial metallicity of $Z=0$ and a hydrogen mass fraction of $X=0.76$. To be concrete, we assume{\footnote{As described in the Introduction, our results do not depend on the choice of an initial NFW profile, but instead hold for any initial profile, even a cored Burkert profile.} that initially baryons and DM distributions can be described with an NFW density profile (\citealp*{NFW}),
\begin{equation}
	\rho_i(r) = \frac{\rhv\,c^3}{3\theta}\frac{1}{r/r_s\,(1+r/r_s)^2} \, ,
	\label{eq:iniNFWprof}
\end{equation}
with the concentration parameter $c$, the definition $\theta \equiv \log(1+c)-c/(1+c)$, the scale radius $r_s = \rv/c$ and the virial radius and density, $\rv$ and $\rhv$, respectively. The latter quantity gives the mean density of a sphere enclosed within $\rv$ and can be calculated using the critical density of the Universe, $\rho_{\rm{crit}}$, at a given redshift $z$:
%
\beq
	\rhv \!=\! \Delta_c \rho_{\rm{crit}}(z) \!=\! \Delta_c \frac{3H^2}{8\pi G} \!=\! \Delta_c\frac{3H_0^2}{8\pi G}\left(\Omega_m(1\!+\!z)^3\!+\!\Omega_\Lambda\right).
	 \label{eq:rhovir}
\eeq
Here, $\Delta_c = 178$ is the standard density contrast at virialisation in the spherical collapse model, $H$ is the Hubble parameter, $H_0$ is the present value of the Hubble parameter, and $\Omega_m$ and $\Omega_\Lambda$ are the present energy densities of matter and the cosmological constant, respectively, in units of the critical energy density.

For the initial NFW profile, we choose a fiducial value of $c=3.5$ for minihalos. However, we note that the properties of DSs stay roughly the same for concentration parameters $c=2-5$, see \citealp*{Ilie11}. We stress again, that the formation of a DS does not depend on the exact form of the initial DM profile; see \autoref{sec:Intro}.

Since the energy generation rate within DSs depends upon cosmological parameters via the DM density (by way of Eq.~(\ref{eq:rhovir}) and Eq.~(\ref{eq:DMheat})), we have to specify a fiducial cosmology. 
For this paper, we took the cosmological parameters from \citealp*{Planck15}, namely a flat $\Lambda$CDM universe with $\(\Omega_m, \Omega_\Lambda, h\) = \(0.308, 0.692, 0.678\)$.

%
\section{Comparison of Model Results: Polytropes, MESA Versions \& Implementations} \label{sec:Comp}
%


We compare the results from different DS implementations and codes to test the robustness of the numerical predictions for the properties of DSs. Polytropic models were studied in F10. MESA was first used by us for DSs in RD15, release 5596. Since then, MESA has been steadily improved and our first results pertaining to this paper were obtained using release 9957. In March 2020, release 12778 came online, with significant improvements e.g. in handling energy conservation or atmospheric boundary conditions. We thus decided to re-do our analysis using release 12778. It is encouraging to see that the global results are very similar between different MESA releases, but it is also true that important details can differ. This is especially critical for the study of pulsations, because the latter depend upon the accuracy of the underlying stellar models. 

We consider two cases: (i) the DM heating is calculated using $\fq = 2/3$ throughout the whole interior of the star, and (ii) the DM heating is set to zero, once the baryonic density drops below a specific density threshold, $\rhot$ (cf. Eq.~(\ref{eq:Heatcut}). Both cases enter in Eq.~(\ref{eq:DMheat}) and Eq.~(\ref{eq:DMheatingLum}). We will refer to cases with DM heating cutoff throughout the text as models ``with heating cutoff'', or simply ``with cutoff'', as opposed to models ``without cutoff''. 

As density threshold for the heating cutoff we consider two different values: $\rhot = 10^{-8} \, \gcmc$ and $10^{-9} \, \gcmc$. 
To visualize which parts of the star are influenced by this choice, the density profile of a DS \textit{without} heating cutoff (i.e. the canonical case with $f_Q = 2/3 = \rm{const.}$) at different times of its evolution from $\sim \! 50 \, \MS$ up to $\sim \! 1000 \, \MS$ is shown in \autoref{fig:Dens-CutOffDev}. 
For the case of a $\sim 50 \, \MS$ DS, the choice of
 threshold of $\rhot = 10^{-8} \, \gcmc$ switches off the DM heating in the outer $\approx 10 \%$ of the star in terms of radius, corresponding to the outer 
 $\approx 5 \%$ of the total stellar mass. In more massive stars, e.g. considering the $\sim 1000 \, \MS$ case, a larger region is affected: the heating cutoff removes the energy input from the outer $20 \%$ of the star (in radial coordinates). Yet, because of the low densities, this region encloses only $\sim 1\%$ of the total stellar mass. For the more realistic heating cutoff at $\rhot = 10^{-9} \, \gcmc$, the stars are even less affected. 

\begin{figure}
	\includegraphics{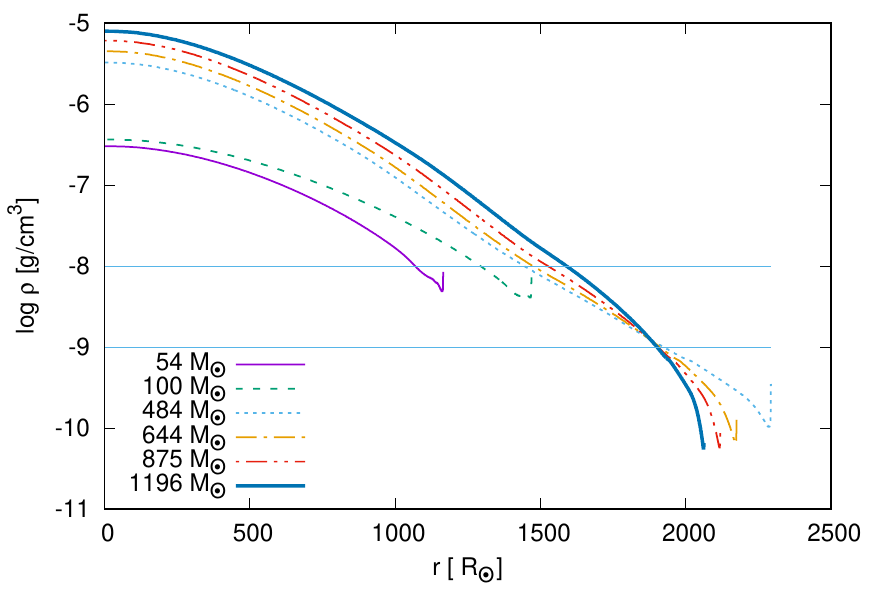}
	\caption{Stellar density in g$/$cm$^3$ as a function of stellar radius in solar radii for a DS \textit{without} DM heating cutoff. Along its evolutionary sequence, the DS grows to progressively higher mass. Here and thoughout this paper, we highlight six timesteps (resp. stellar masses): $54 \, \MS$ (purple thin solid), $100 \, \MS$ (green dash), $484 \, \MS$ (light-blue dot), $644 \, \MS$ (orange dash-dot), $875 \, \MS$ (red dash-dot-dot), and $1196 \, \MS$ (blue thick solid). The two horizontal lines mark the two considered density thresholds for the heating cutoff at $\rhot = 10^{-8}\, \gcmc$ and $10^{-9} \, \gcmc$, respectively. (Models with DM heating cutoff implemented, in which we ignore DM heating from regions where the gas density is too low, are not shown in this plot, but the density profiles are essentially the same.) In all models, -with or without DM heating cutoff-, the stellar density increases close to the stellar surface, as a result of ``density inversion'', see \autoref{sec:Stab_MassLoss}.}
	\label{fig:Dens-CutOffDev}
\end{figure}


Quantitatively, the deviations in stellar properties of models with heating cutoff do not exceed $\sim 1 \%$, compared to models with the canonical case $f_Q=2/3=\rm{const}$. This is true during the entire evolution up to $\sim \, 1000 \, \MS$. This shows that the exact treatment of the functional shape of $\fq$ in the outer layers of the star is not of importance for the star as a whole: the largest fraction of energy is produced in the core, where the DM and baryon densities are highest. Removing the minor contribution from DM heating in the outer layers does not give rise to significant changes in the stellar properties. 


This conclusion is confirmed by the comparison shown in \autoref{tab:comp}, where we compare the outcomes from polytropes to different MESA models, for global stellar quantities, such as luminosity, radius, effective and central temperature, and central density. For any given stellar property, the differences between models with and without heating cutoff within the same MESA release are much smaller than the differences between results from different MESA releases.  
The predictions between MESA 5596 and the newer ones MESA 9957 and now MESA 12778 agree within $\approx 15 \%$ for all considered stellar masses. These deviations arise mostly from the improved implementation of the derivatives of the DM heating rate. While the differences between MESA and polytropes are most noticeable, they are still within factors of a few. For a detailed comparison between the polytropes and the first MESA models, we refer to Section 4 of RD15.

In the \lp~of \autoref{fig:RadLum}, we show the evolution of stellar radii $R_{\star}$ and effective temperature $T_{\rm eff}$, respectively, as the DS grows in mass, for models with heating cutoff at $\rhot = 10^{-9}\, \gcmc$ and those without cutoff: the respective curves are so similar that they lie on top of each other. We can see that 
$\Teff$ stays roughly constant at $\sim 4000$ K before it increases sharply at $\sim 200~ \MS$. Around the same time, the growth in radius stalls, the DS shrinks a bit, before the monotonic growth continues past $\sim 1000 ~\MS$. This behaviour can be also read off in \autoref{tab:comp}. This phenomenon will become clearer when we describe the dynamics in \autoref{sec:Stab_MassLoss}.
The \rp~of the figure shows the evolution of the stellar luminosity $L_{\star}$, again for models with and without heating cutoff, and their curves lie almost on top of each other. $L_{\star}$ grows roughly linearly with the mass of the star $\Ms$, which is a typical feature for massive stars.

\begin{figure*}
     \begin{minipage}[b]{0.5\linewidth}
      \centering\includegraphics[width=8cm]{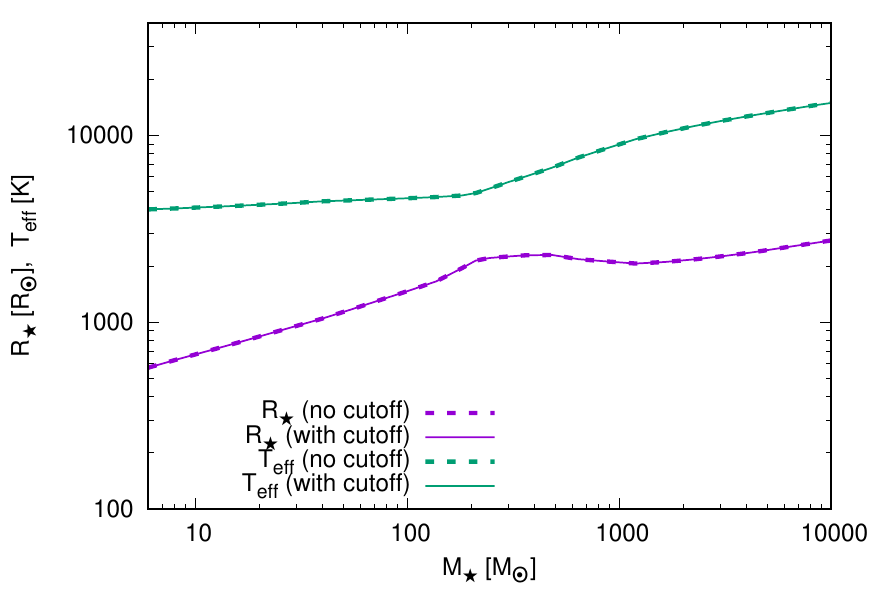}
     \hspace{0.1cm}
    \end{minipage}%
 \begin{minipage}[b]{0.5\linewidth}
      \centering\includegraphics[width=8cm]{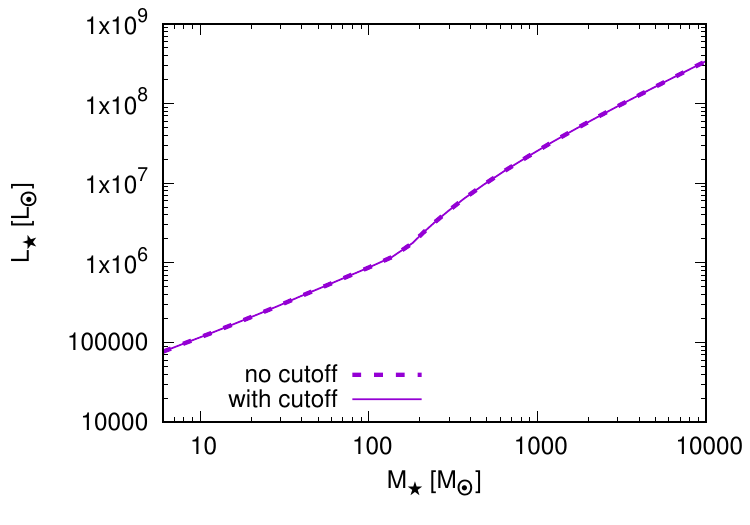}
     \hspace{0.1cm}
    \end{minipage}
 \caption{Evolution of global stellar properties: \Lp: stellar radius (lower curves) and effective temperature (upper curves) with increasing DS mass. DSs with masses above around $500~ M_{\odot}$ shrink before the monotonic growth in radius continues above $1000~ M_{\odot}$. The effective temperature remains almost constant below about $200 ~M_{\odot}$, before it starts increasing monotonically.
	\Rp: Stellar luminosity with increasing DS mass.  Notice that the respective curves for ``no cutoff'' (dash curves) and ``with cutoff'' (solid curves) are so similar that they lie basically on top of each other in each panel.}
 \label{fig:RadLum}
\end{figure*}

\begin{table*}
 \begin{minipage}{\linewidth}
 \centering
	\begin{tabular}{cc|c|c|c|c|c|c}
\hline
Models   &   & $M_{\star}$   & $L_{\star}$ 			& $R_{\star}$ 	& $T_{\rm{eff}}$	& $T_c$       		& $\rho_c$ \\
    	 & 	 & $[M_{\odot}]$ & $[10^6~ L_{\odot}]$ 	& $[\RS]$ & $[10^3~ \rm{K}]$ 	& $[10^5~\rm{K}]$ 	& [$10^{-6}\,\gcmc$] \\      
\hline \hline
MESA 12778& $\fq = 2/3$				& 100	& 0.88 	& 1469 	& 4.6 	& 2.6	& 0.36 	\\ 
MESA 12778& $\rhot = 10^{-9}\,\gcmc$	& 100	& 0.88 	& 1469 	& 4.6	& 2.6	& 0.36 	\\
MESA 12778& $\rhot = 10^{-8}\,\gcmc$   & 100 	& 0.88 	& 1469 	& 4.6	& 2.6	& 0.36	\\
\hline
MESA 9957& $\fq = 2/3$				& 100	& 0.76 	& 1513 	& 4.4 	& 2.5	& 0.34 	\\ 
MESA 9957& $\rhot = 10^{-9}\,\gcmc$	& 100	& 0.76 	& 1513 	& 4.4	& 2.5	& 0.34 	\\
MESA 9957& $\rhot = 10^{-8}\,\gcmc$ & 100 	& 0.75 	& 1504 	& 4.4	& 2.5	& 0.34	\\
\hline
MESA 5596 & from RD15 				& 100 	& 0.76 	& 1493 	& 4.4 	& 2.6 	& 0.37 	\\ 
\hline
Polytrope & from F10 				& 100	& 1.2 	& 1120  & 5.7  	& 3.4 	& 0.74 	\\ 
\hline  \hline 
\\
MESA 12778& $\fq = 2/3$			  & 500	& 9.7	& 2291 	& 6.7	& 7.6	& 3.3 	\\ 
MESA 12778& $\rhot = 10^{-9}\,\gcmc$ & 500	& 9.7	& 2291	& 6.7 	& 7.6	& 3.3 	\\
MESA 12778& $\rhot = 10^{-8}\,\gcmc$ & 500	& 9.7 	& 2291 	& 6.7	& 7.6	& 3.3 	\\
\hline
MESA 9957& $\fq = 2/3$				& 500	& 10.4	& 2304 	& 6.8	& 8.4	& 4.3 	\\ 
MESA 9957& $\rhot = 10^{-9}\,\gcmc$ & 500	& 10.3	& 2300	& 6.8 	& 8.4	& 4.3 	\\
MESA 9957& $\rhot = 10^{-8}\,\gcmc$ & 500	& 10.2 	& 2282 	& 6.8	& 8.4	& 4.3 	\\
\hline
MESA 5596& from RD15  				& 500	& 10.4	& 2257	& 6.9 	& 8.6	& 4.6 	\\ 
\hline
Polytrope& from F10 				& 500	& 9.7 	& 2000 	& 7.2 	& 8.3 	& 4.3 	\\ 
\hline \hline
\\
MESA 12778& $\fq = 2/3$ 				& $10^3$& 31.8	& 2061 	& 9.5	& 12.3 	& 8.0 	\\ 
MESA 12778& $\rhot = 10^{-9}\,\gcmc$   & $10^3$& 31.8	& 2061 	& 9.5	& 12.3	& 8.0 	\\
MESA 12778& $\rhot = 10^{-8}\,\gcmc$   & $10^3$& 31.8	& 2061 	& 9.5 	& 12.3	& 8.0 	\\
\hline
MESA 9957& $\fq = 2/3$ 				& $10^3$& 25.9	& 2851 	& 7.7	& 12.1 	& 8.5 	\\ 
MESA 9957& $\rhot = 10^{-9}\,\gcmc$ & $10^3$& 25.7	& 2840 	& 7.7	& 12.1	& 8.5 	\\
MESA 9957& $\rhot = 10^{-8}\,\gcmc$ & $10^3$& 25.4	& 2786 	& 7.8 	& 12.1	& 8.5 	\\
\hline
MESA 5596& from RD15				& $10^3$& 25.9 	& 2437 	& 8.3 	& 12.4 	& 9.0 	\\ 
\hline
Polytrope& from F10 				& $10^3$& 17	& 2580 	& 7.5 	& 9.8 	& 4.6 	\\ 
\hline \hline 
\end{tabular}
\end{minipage}
\caption{Comparison of properties of DSs along their evolutionary sequence for fixed $m_{\chi}=100$ GeV and $\dot M = 10^{-3} M_{\odot}/$yr (''SMH100'') between different MESA models and polytropes. We compare global quantities, such as stellar luminosity $L_{\star}$, radius $R_{\star}$, effective temperature $T_{\rm{eff}}$, central temperature $T_c$ and total central density 
$\rho_c$ (i.e. gas and DM density) for stars at fixed masses $M_{\star}$. Notice that the differences between models with and without heating cutoff within the same MESA release are smaller than the results between different MESA releases, or polytrope models. Remark: in MESA 12778, the star contracts between around $500-1000 ~M_{\odot}$ (see also \autoref{fig:Dens-CutOffDev}), resulting in a higher $T_{\rm{eff}}$ compared to other models in this mass range. More explanations can be found in the text.}
\label{tab:comp}
\end{table*}

To sum up, the comparison shows that the stellar properties depend only marginally on the exact value or functional dependence of $f_Q$. Therefore, we conclude that the approximation $f_Q(\rhos)= \rm{const.}$, that was applied in the previous DS papers is justified, assuming that the effects of DM annihilation products in the outer regions of the star do not prevent DSs from growing beyond a certain mass (see discussion in \autoref{sec:Intro}). In the rest of this work, we will show results for the model with DM heating cutoff at 
$\rhot = 10^{-9}\,\gcmc$. However, we repeated all steps in the analysis with the other two considered models (no heating cutoff and $\rhot = 10^{-8}\,\gcmc$) and did not find any changes between the different treatments that would affect our conclusions.

%
\section{Dynamical Stability, Eddington Limit and Mass Loss} \label{sec:Stab_MassLoss}
%

Previous studies have shown that DSs become more and more radiation pressure-dominated as they grow in mass (F10, RD15), and it was shown that lower-mass stars are well approximated by $(n=3/2)$-polytropes, while higher-mass stars are well approximated by $(n=3)$-polytropes. 
Supermassive DSs acquire extended, extremely diluted, weakly convective envelopes, causing the typical features of ``superadiabaticity''. We confirm this trend for the most massive models investigated here.
 
\autoref{fig:ProfQuant} (top left) shows the ratio of gas pressure to total pressure, in the literature known as $\beta=P_{\rm{gas}}/P$. Small values of $\beta$ indicate radiation pressure-dominated interiors. In fact, this is the case above several hundred solar masses, although $\beta$ increases always close to the stellar surface, even for stars of smaller mass.  
The top right of \autoref{fig:ProfQuant} shows the corresponding run of specific entropy within the DSs. Regions of efficient convection are indicated by layers of constant entropy. All of our stellar models have convective cores, which we also checked by looking at the sign of the Brunt-V\"ais\"al\"a frequency squared. Layers whose entropy declines with radius are convectively unstable; this is true close to the stellar surface, where the convective heat transport is particularly inefficient. We can see that low-mass DSs are almost entirely convective, except close to the stellar surface. As the mass of the DS grows, layers develop within which the entropy increases. This indicates that a radiative region develops between the convective core and the surface layers.

The development of such radiation pressure-dominated, weakly convective envelopes in growing DSs presents a potential problem, if those envelopes cause the whole star to become dynamically unstable. In this section, we therefore investigate the effects that could prevent DSs from growing supermassive: dynamical instabilities and mass loss possibly caused by super-Eddington winds. 

A star is dynamically unstable, if it collapses upon a small, initial compression. This is the case if the weight increase arising from the compression is larger than the increase of the pressure. The criterion for dynamical instability can be derived with the adiabatic exponent, $\Gamma_1$, which quantifies how the gas reacts to compressions. The condition for global 
instability is given by
\beq
	\int \left(\Gamma_1 - \frac{4}{3}\right) \frac{P}{\rhos} \, {\rm d}\,  m \, < 0,
	\label{eq:dynInst}
\eeq
where the integral runs over the whole volume of the star. The adiabatic exponent is given by $\Gamma_1 \equiv \left( \frac{ \rm{d \, ln}P }{\rm{d \, ln} \rhos}\right)_{\rm ad}$, $P$ is the pressure and $\rhos$ the stellar density. This means that even though the adiabatic exponent $\Gamma_1$ can locally drop below the critical value of $4/3$, the overall stability of the star is assured as long as the integral in Eq.~\eqref{eq:dynInst} is positive.

The adiabatic exponent $\Gamma_1$ usually falls below the value of $4/3$ at the surface layers of DSs. Since the densities in these regions are particularly low, this could enhance the factor of $P/\rhos$ in the outer regions and lead to an instability. We check for the global stability of the star by calculating the integral in Eq.~(\ref{eq:dynInst}) numerically, for all evolutionary stages of each DS model we considered. We find that the condition for global dynamical instability is never met, that is, the integral is always positive. This holds for all models, i.e.\! with and without DM heating cutoff.

Next, we study the so-called Eddington limit.
This limit refers to a condition in the star where the outward acceleration due to radiation pressure balances the inward force due to gravity, in hydrostatic equilibrium. If this limit is surpassed at the surface, it is believed that mass outflows should arise. The classical Eddington factor is defined as
\begin{equation} \label{eq:Edd}
 \frac{L}{L_{\rm{Edd}}} = \frac{g_{\rm{rad}}}{g} = \frac{\kappa_e L}{4\pi c~ GM},
\end{equation}
where $L$, $M$ and $\kappa_e$ are the (total) luminosity, mass and electron-scattering opacity, respectively. $g=GM/r^2$ is the gravitational acceleration and $g_{\rm{rad}} = \kappa_e L/4\pi r^2$ is the radiative acceleration due to electron scattering opacity. Hence, exceeding the Eddington limit is usually refered to $L/L_{\rm{Edd}} > 1$.
While this condition provides a sufficient instability criterion to stars, it is not a necessary one. When Eq.(\ref{eq:Edd}) is applied inside the star, as we will discuss shortly, we must replace $L$ on the left-hand side with the radiative luminosity $L_{\rm{rad}}$. The true opacity can significantly exceed $\kappa_e$: when the opacity is large, convection kicks in and $L_{\rm{rad}}$ is reduced. As a result, the Eddington limit is not reached, and the star is safe again against instability.  In fact, even an excess of $>1$ of the \textit{local} Eddington factor, which varies within the stellar interior, does not by itself necessarily indicate an instability, as already shown in \citealp*{Joss73}.
We follow recent literature, especially \citealp*{Sanyal15}, and thus consider a more plausible definition of the local Eddington factor, according to 
\begin{equation}
 \frac{L_{\rm{rad}}(r)}{L_{\rm{Edd}}(r)} = \frac{\kappa(r) (L(r)-L_{\rm{conv}}(r))}{4\pi c~ GM(r)}.
 \label{eq:eddington_lim}
\end{equation}
In the rest of this paper, we refer to this definition for the Eddington factor, and the Eddington limit is reached, once  $L_{\rm{rad}}(r_l)/L_{\rm{Edd}}(r_l)=1$ for some radial coordinate $r_l \leq R_{\star}$.
Now, the analysis in \citealp*{Sanyal15} has proved as a useful comparison to our results, because i) it is a recent study, ii) it considers models for very massive main-sequence stars up to $500 ~M_{\odot}$ (i.e. similar to our mass range), and iii) we find the interpretation of the physical effects described there (e.g. density inversion close to the surface) very plausible in explaining our own results, regarding super-Eddington layers. 
A super-Eddington layer can form, especially very close to the stellar surface where convective energy transport is highly inefficient, pushing $L_{\rm{rad}}(r)/L_{\rm{Edd}}(r)$ close to, or above one. However,
it turns out that such a super-Eddington layer inside a star does not necessarily lead to a departure from hydrostatic equilibrium or to any mass outflow, even if the maximum of Eq.(\ref{eq:eddington_lim}) -$\max(L_{\rm rad}(r)/L_{\rm Edd}(r))_{r\leq R_{\star}}$- is often attained very close to the stellar surface. The local Eddington factor Eq.~\eqref{eq:eddington_lim} is shown for our models in \autoref{fig:ProfQuant} (middle left).
For instance, \citealp*{Sanyal15} find that their models remain stable, even if $\max(L_{\rm rad}/L_{\rm Edd}) \sim 7$. Our models exemplify smaller maximum values, 
$\max(L_{\rm rad}/L_{\rm Edd}) < 1.8$, i.e. only slightly overshoot the limit for some range in stellar mass during the evolution,  as can be seen in \autoref{fig:ProfQuant} (middle right).
Nevertheless, our MESA runs do not exhibit difficulties across this stellar mass range, where the overshooting happens.
Instead, we observe a similar behaviour than reported in \citealp*{Sanyal15}: the star counteracts a super-Eddington luminosity by developing a positive density gradient in a thin shell very close to the stellar surface, i.e. a ``density inversion'' (compare to \autoref{fig:Dens-CutOffDev}).

%
%

This discussion on Eddington limits naturally leads up to the discussion of opacity. 
 The opacity curves, along with the run of the neutral fraction of H and He, are shown in \autoref{fig:ProfQuant} (bottom left and right panels, respectively): for low values of $T \lesssim 10^4$ K, the opacity is given by the absorption of H$^-$. The peak at $T \simeq 10^4$ K is due to the bound-free absorption of H and H$^{-}$ ("hydrogen bump"). For higher values of the temperature, the opacity follows Kramer's law, with the smaller peaks at higher temperature being due to the first and second ionization of He, respectively ("helium bumps"). At even higher temperatures, the opacity reaches a constant value due to Thomson e$^{-}$-scattering, which is temperature-independent once the stellar material is fully ionized, as can be also seen in \autoref{fig:ProfQuant} (bottom right), where the transition from neutral to ionized layers of H (at lower $T$) and He (at higher $T$) is shown. 
 As DSs grow and become hotter, the opacity bumps become less pronounced. Moreover, the hydrogen bump moves closer to the surface for higher mass DSs, which, in turn, results in an increase in the Eddington factor Eq.~\eqref{eq:eddington_lim}, see middle left in \autoref{fig:ProfQuant}. In fact, the maximum Eddington factors along the evolutionary sequence, which we show in \autoref{fig:ProfQuant} (middle right), occur close to the stellar surface of our models, and they can be connected to the opacity peak of H. In general, whenever one of such opacity peaks is situated sufficiently close to the stellar photosphere, the densities in these layers are so small that convective energy transport becomes inefficient; super-Eddington layers develop which are stabilized by a positive (i.e. inward-directed) gradient in density and gas pressure.

We note that we found no difference between models with or without DM heating cutoff with respect to the development of convective zones, the shape of the opacity curve, or the values for the Eddington factors.

\begin{figure*} 
\begin{minipage}{0.5\linewidth}
     \centering\includegraphics[width=8cm]{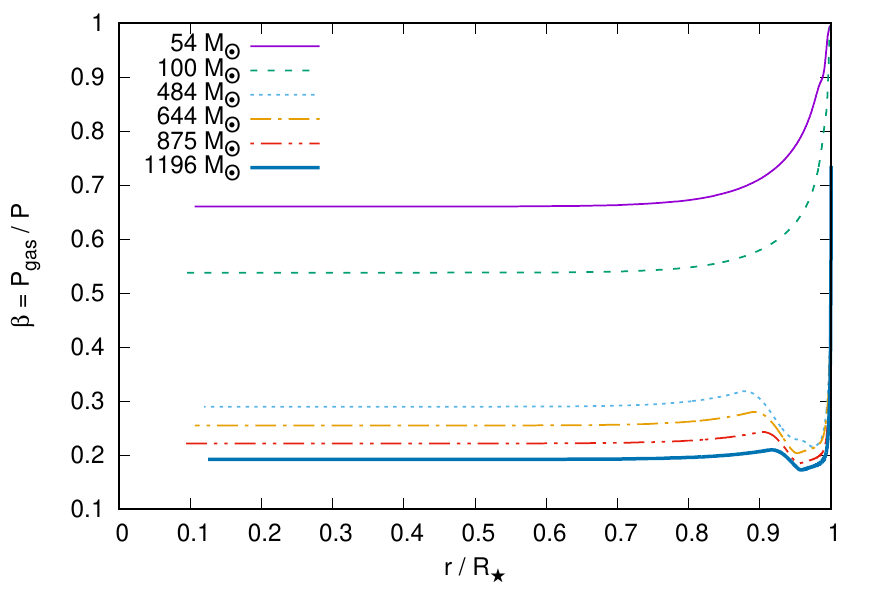}
     \vspace{0.05cm}
    \end{minipage}%
    \begin{minipage}{0.5\linewidth}
      \centering\includegraphics[width=8cm]{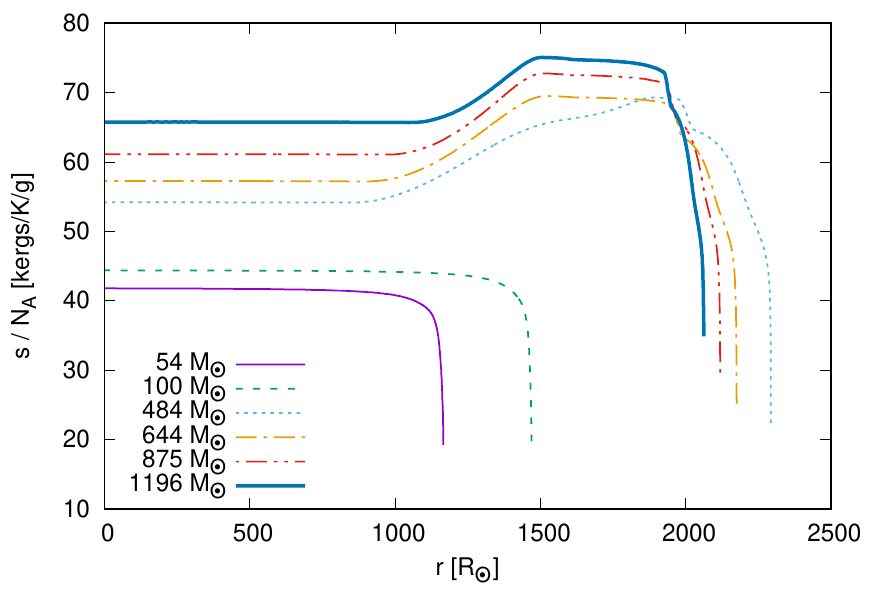}
     \hspace{0.05cm}
    \end{minipage}
 \begin{minipage}{0.5\linewidth}
     \centering\includegraphics[width=8cm]{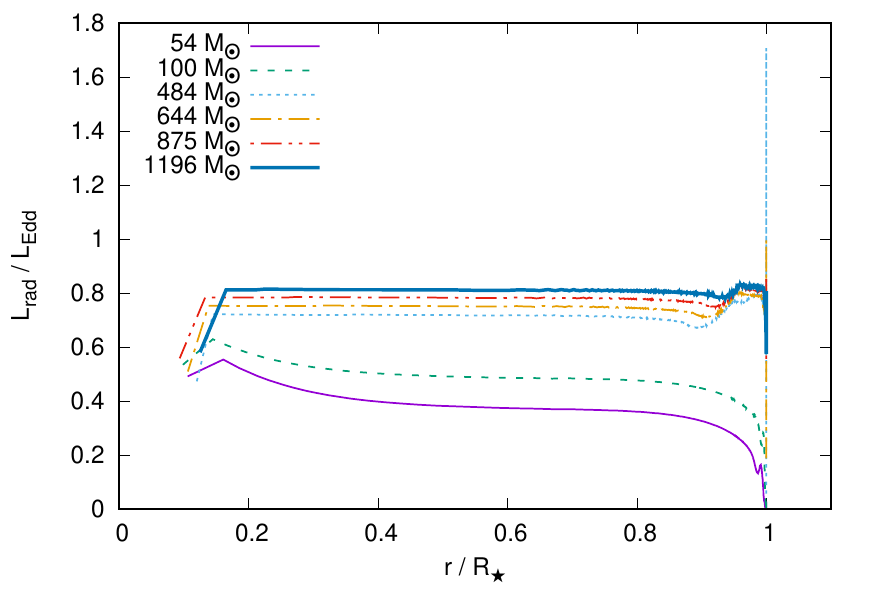}
     \vspace{0.05cm}
    \end{minipage}%
    \begin{minipage}{0.5\linewidth}
      \centering\includegraphics[width=8cm]{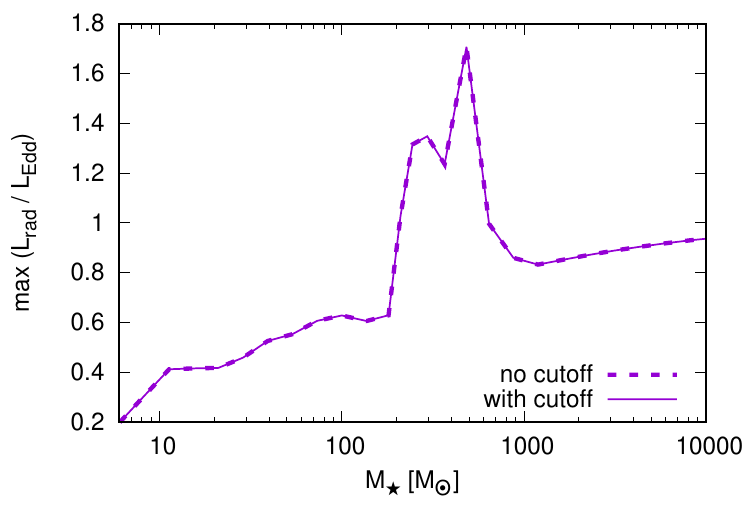}
     \hspace{0.05cm}
    \end{minipage}
    \begin{minipage}{0.5\linewidth}
      \centering\includegraphics[width=8cm]{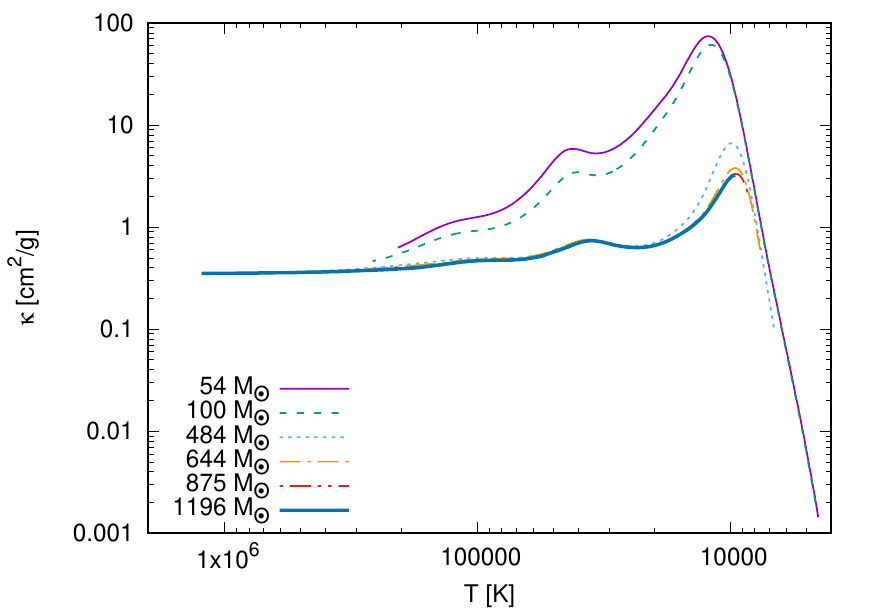}
     \hspace{0.05cm}
    \end{minipage}%
    \begin{minipage}{0.5\linewidth}
      \centering\includegraphics[width=8cm]{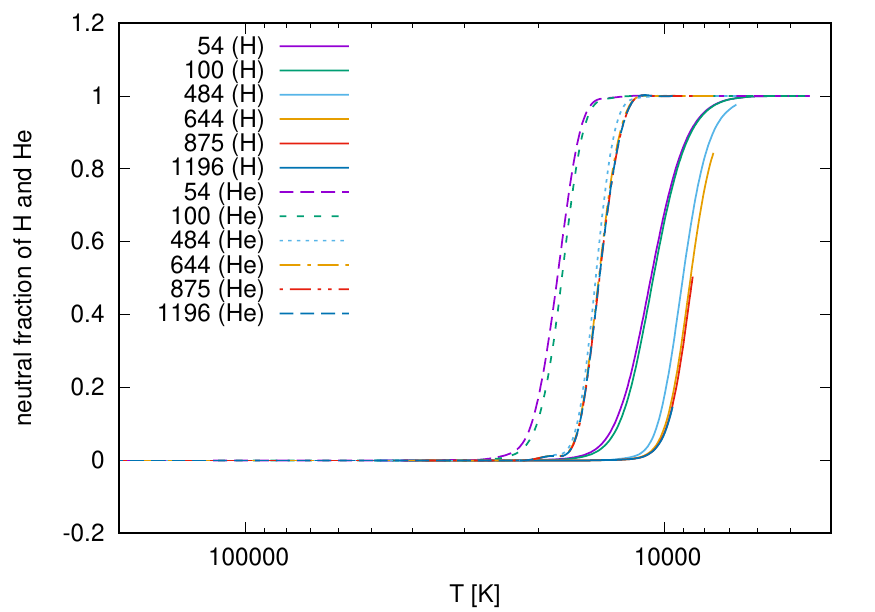}
     \hspace{0.05cm}
    \end{minipage}
 \caption{\emph{Top left}: Ratio of gas pressure to total pressure $\beta$ as a function of relative stellar radius, $r/R_\star$, for DSs of different masses as indicated in the legend (the labels for the stellar masses are the same as in the caption of \autoref{fig:Dens-CutOffDev}). 
 \emph{Top right}: Specific entropy as a function of stellar radius, same labels as in \autoref{fig:Dens-CutOffDev}.
 \emph{Middle left}: Eddington factor, $L_{\rm rad}/L_{\rm Edd}$ as a function of relative stellar radius, $r/R_\star$; same labels as in \autoref{fig:Dens-CutOffDev}. Notice that $L_{\rm rad}/L_{\rm Edd} > 1$ at $484 ~M_{\odot}$. \emph{Middle right}: Maximum values of the Eddington factor, $\rm{max}(L_{\rm{rad}}/L_{\rm{Edd}})$, throughout the entire evolution of the DS. The factor slightly exceeds 1 at stellar masses between $\sim 200 - 600 ~M_{\odot}$; the maximum value is always attained close to the surface of the respective models. For comparison, this plot shows the result with (solid) and without (dash) DM heating cutoff; the curves lie almost on top of each other.
 \emph{Bottom left}: Opacities as a function of temperature for our stellar models; same labels as in \autoref{fig:Dens-CutOffDev}. \emph{Bottom right}: Plot of the run of neutral fraction of H (solid curves) and He (dashed curves), respectively, each for our depicted stellar models.} \label{fig:ProfQuant}
\end{figure*}

%
\section{Non-adiabatic Pulsations} \label{sec:Puls}
%

Pulsations in supermassive DSs would provide a key mechanism for
distinguishing them from early galaxies in observations. However, if
they reach an amplitude sufficient to trigger mass loss, they could
also potentially halt the mass growth of DSs via accretion. With these
two motivations in mind, we turn in this section to the question of
whether pulsations are expected to self-excite in DSs. We stress that
uncertainties in the stellar models translate into uncertainties in
our pulsation calculations. Therefore, a further understanding of the
former will be required in order to settle definitely the question of
whether or which pulsation excitation mechanisms might be at play in
DSs.

In RD15, the periods of radial (i.e. $l=0$) pulsation modes with different overtone number $n$, where $n=1$ is the fundamental mode, and higher overtone modes with $n>1$, were modeled using the ADIPLS stellar oscillation
code of \citealp*{adipls}, built into MESA. It adopts the adiabatic
approximation whereby the transfer of heat between neighboring fluid
elements is neglected. While this approximation allows considerable
simplification of the analysis, it has the drawback that information
about wave excitation and damping is lost; as a result the global
stability or instability of modes cannot be determined. In the present
paper, we address this limitation by using the GYRE stellar
oscillation code (\citealp*{Gyre, TGZ18}) to calculate \emph{non-adiabatic} eigenfrequencies,
eigenfunctions and accompanying properties of our DS models. To ensure
that no modes are missed, we employ the contour method (\citealp*{GT20}) implemented in release 6.0 of GYRE.

\subsection{Stability Calculations}

For the sequence of DS models spanning the mass range $10\,M_{\odot}
\leq M_{\ast} \leq 10^{4}\,M_{\odot}$, we use GYRE's contour method to
find non-adiabatic eigenfrequencies $\sigma$ of radial ($l=0$)
modes.  These are complex quantities; with an assumed time dependence
$\propto \exp(-{\rm i}\sigma t)$ for perturbations, the real part of
the eigenfrequency $\sigma_{\rm R}$ determines the linear frequency
$\nu_{\rm osc} = \sigma_{\rm R}/(2\pi)$ of a mode. The corresponding
imaginary part indicates whether the mode is globally unstable
($\sigma_{\rm I} > 0$) or stable ($\sigma_{\rm I} < 0$). Unstable
modes grow exponentially with time from an initial infinitesimal
perturbation, with an $e$-folding time $\tau_{\rm osc} = 1/\sigma_{\rm
  I}$. This growth tapers off when their amplitude is sufficiently
large that non-linear effects become important. Because GYRE is a
linear code, we cannot model this non-linear saturation, nor predict
the final amplitude of unstable modes.

\begin{figure} 
  \centering
  \includegraphics{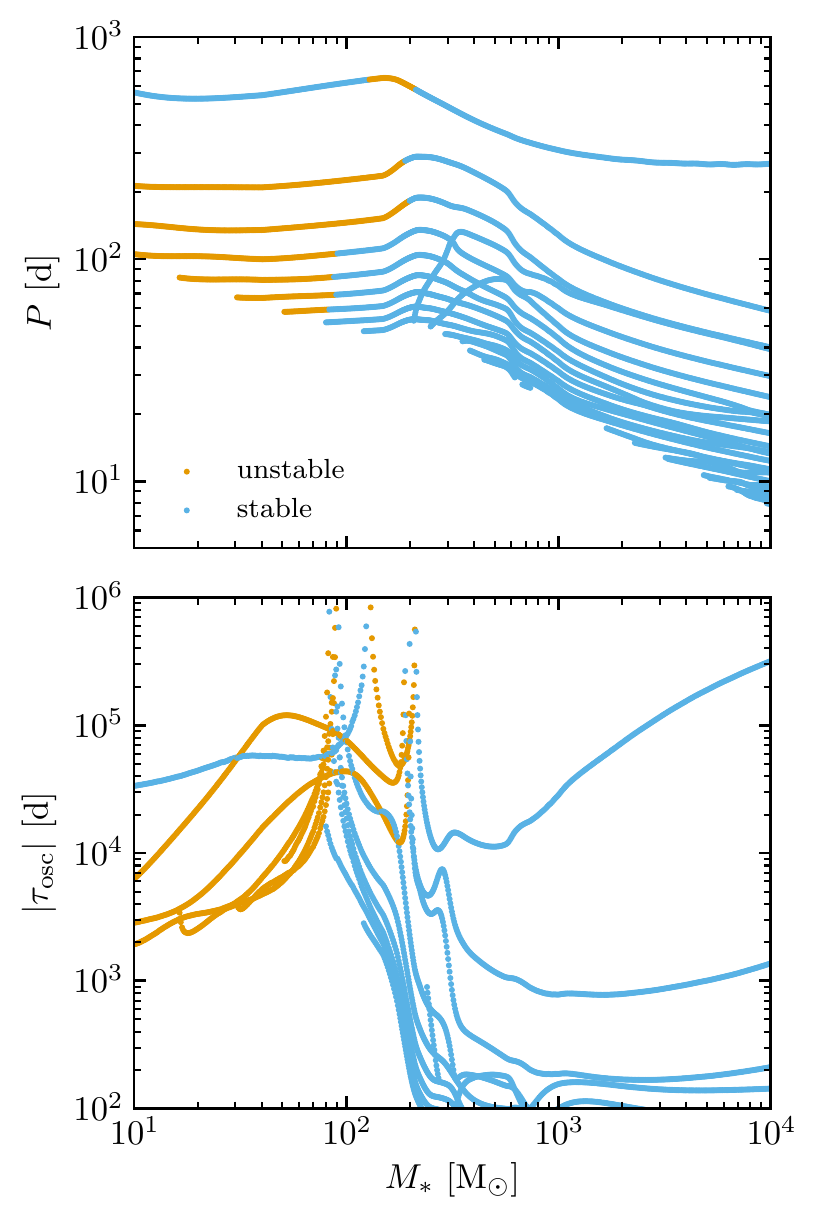}
  \caption{Modal diagram for our DS models, plotting the mode period
    $P$ in days (upper panel) and growth/damping timescale $|\tau_{\rm osc}|$ in days
    (lower panel), both as a function of stellar mass $M_{\ast}$. Unstable
    (stable) modes are indicated using orange (blue)
    markers. Notice that the periods and timescales are in the rest frame of the star.} \label{f:modal-diagram}
\end{figure}

\autoref{f:modal-diagram} displays the results from our calculations
in the form of a \emph{modal diagram}, showing how the
eigenfrequencies evolve as the DS mass increases. The upper panel
plots the mode periods $P \equiv 2\pi/\sigma_{\rm R}$, and the lower
panel the modulus of the growth/damping $e$-folding timescale $|\tau_{\rm osc}|$,
both as a function of $M_{\ast}$.  Modes that are unstable (stable)
are indicated using orange (blue) markers. Note that periods and
timescales in the observer's rest frame are a factor 21 larger,
because they are redshifted by a factor $(1 + z_{\ast}) \sim 21$,
where $z_{\ast}$ is the redshift of the star%
\footnote{The most massive and oldest star considered here is at a
  redshift of $z_{\ast} = 19.84$; since the error of an exact
  treatment is less than 1\%, we can safely ignore the change of the
  redshift of the star in the mass range that we consider and
  approximate $z_{\ast} \sim 20$.}%
.

In the upper panel, the longest-period mode is the fundamental mode;
higher overtones occur toward progressively shorter periods, down to a
limit set by the acoustic cutoff frequency of the stellar
atmosphere. At frequencies above this cutoff, outward-propagating
waves cannot be reflected at the atmosphere, and no standing waves can
form in the star. 

As the mass increases, the mode periods remain approximately constant
up to $M \approx 200\,M_{\odot}$, and then decrease steadily with further
growth. Broadly, this behaviour is a consequence of the radius
evolution shown in the \textit{left panel} of \autoref{fig:RadLum}. At lower masses, the
radius grows approximately as $R_{\ast} \sim M_{\ast}^{1/3}$. Because
radial-mode periods scale proportionally with a star's dynamical
timescale $(R_{\ast}^{3}/GM_{\ast})^{1/2}$, the mass and radius
changes balance and the period evolution remains flat. At higher
masses, however, the radius plateaus and so the shortening dynamical
timescale drives mode periods down. Similar behaviour can be seen in RD15
(cf. the mid-left panel of their Figure 7).

The figure reveals that all modes apart from the fundamental are
unstable at the low-mass limit. Toward higher stellar masses, this instability
gradually morphs into stability, with the highest-overtone
(shortest-period) modes transitioning first. Before the overtone
instability completely vanishes, the fundamental mode enters a phase
of instability, spanning the mass range $130\,M_{\odot} \la M_{\ast}
\la 210\,M_{\odot}$. Once that phase is over, all modes remain stable
through to the upper mass limit.

\subsection{Excitation Mechanism}

To explore the excitation mechanism at work in the unstable modes
plotted in \autoref{f:modal-diagram}, we focus on a DS model with
mass $M_{\ast} = 100\,M_{\odot}$, in which the fundamental mode ($n=1$) is
stable, but the first- and second-overtone modes ($n=2$ and $n=3$, respectively) are
unstable. \autoref{f:eig-func} plots the radial displacement
perturbation $\delta r$ as a function of relative stellar radius $r$ for
these three modes%
\footnote{Strictly speaking, $\delta r$ is a complex quantity, and we
  plot only the real part in the figure; however, the corresponding
  imaginary parts remain small throughout the star for the three modes
  shown.}. The morphologies seen in the figure are typical to radial
modes: the fundamental mode has one node (one zero) only at the stellar centre,
while the first-overtone mode has two nodes, and the second-overtone
mode three.

\begin{figure} 
  \centering
  \includegraphics{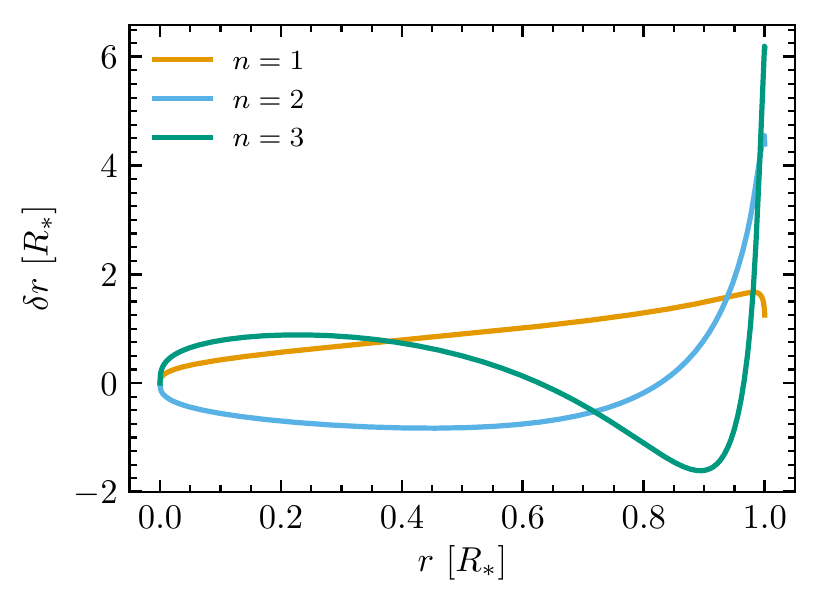}
  \caption{Radial displacement perturbations $\delta r$ for the
    fundamental (F), first-overtone (1O) and second-overtone (2O)
    modes (in orange, blue, green, resp.), plotted against relative stellar radius $r$ for the DS model with $M_{\ast} = 100\,M_{\odot}$. Because GYRE is a linear code, the
    normalization of these eigenfunctions is arbitrary; by convention,
    GYRE scales them so that the mode inertia (see \citealp*{Aerts}, their equation 3.139) equals
    $M_{\ast} R_{\ast}^{2}$.}
    \label{f:eig-func}
\end{figure}

\begin{figure} 
  \centering
  \includegraphics{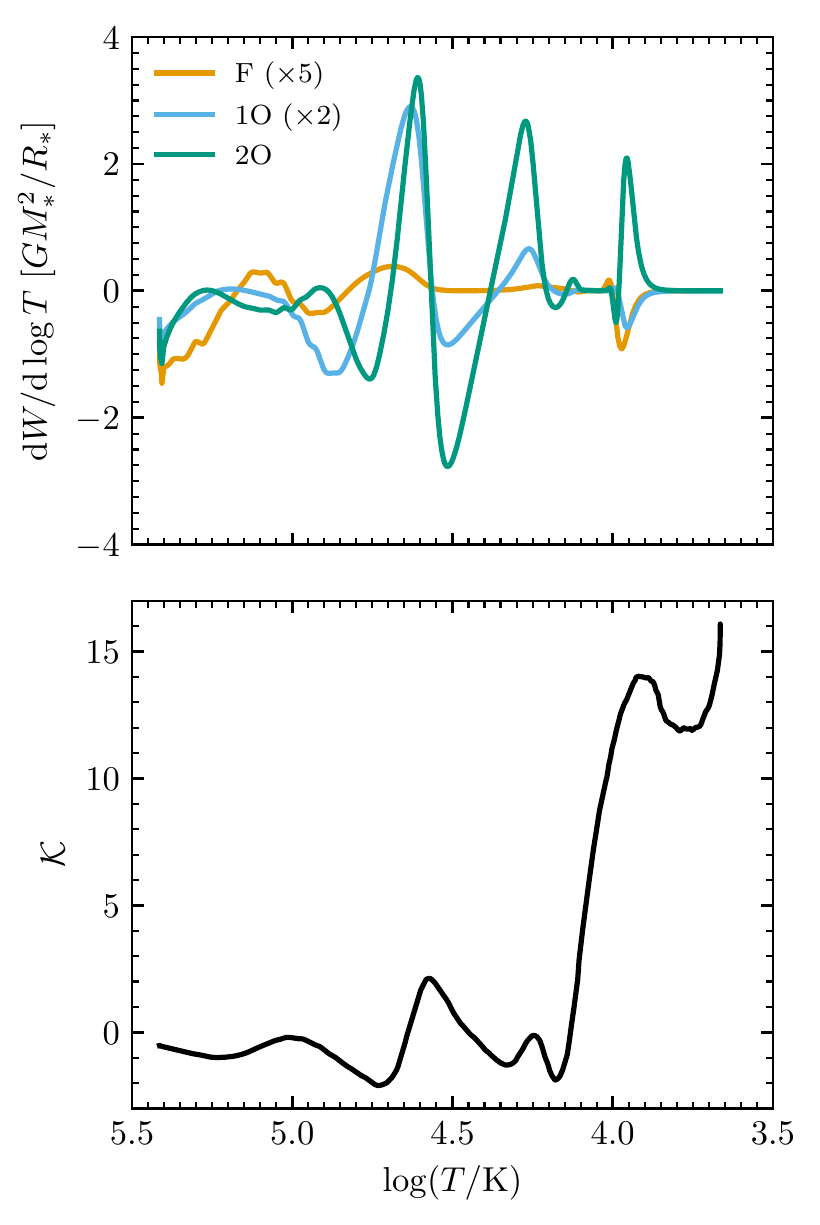}
  \caption{Differential work functions for the fundamental (F),
    first-overtone (1O) and second-overtone (2O) modes (upper panel),
    and the quantity $\mathcal{K}$ defined in
    equation~(\ref{e:kappa-gamma}) (lower panel), plotted against
    $\log T$ for the $M_{\ast} = 100\,M_{\odot}$ DS model. Notice that in the upper panel, the results for F and 1O have been multiplied by factors $5$ and $2$, respectively, for better visualization.}
    \label{f:work-func}
\end{figure}

For the same model and the same three modes, the upper panel of
\autoref{f:work-func} plots the differential work ${\rm d}W/{\rm
  d}\log T$ as a function of $\log T$ (we chose temperature as the
abscissa to better display the driving and damping). The differential
work quantifies the change in mode energy over one cycle, per unit
decade in temperature within the star; it is one of the standard
outputs of GYRE. Regions of the star where the differential work is positive
(negative) correspond to locations where the mode is driven
(damped). Integrating the differential work over the entire star leads
to the total work
\begin{equation}
  W = \int_{\log T_{\rm s}}^{\log T_{\rm c}} \frac{{\rm d}W}{{\rm d}\log T} \, {\rm d}\log T,
\end{equation}
where $T_{\rm s}$ and $T_{\rm c}$ are the surface and central
temperature, respectively, of the star. If $W > 0$ then the mode is
globally unstable and will grow over time; conversely, $W < 0$
indicates a globally damped mode.

The data plotted in the upper panel of \autoref{f:work-func} reveal
strong driving of the second-overtone mode at three distinct locations
in the stellar envelope, corresponding to peaks in the differential
work at $\log T \approx 4.6$, $\log T \approx 4.3$, and $\log T
\approx 3.95$. For the first-overtone mode, only the two inner
(hotter) locations are active, and for the fundamental mode only the
innermost. In the latter case, the driving is not able to overcome
damping elsewhere in the star, and the mode is globally stable.

To explore the origins of these driving regions, we consider the
quantity
\begin{equation} \label{e:kappa-gamma}
  \mathcal{K} = \kappa_{T} + \frac{\kappa_{\rho}}{\Gamma_{3} - 1},
\end{equation}
where
\begin{gather}
  \kappa_{T} \equiv \left( \frac{\partial\ln\kappa}{\partial\ln T} \right)_{\rho},
  \qquad
  \kappa_{\rho} \equiv \left( \frac{\partial\ln\kappa}{\partial\ln \rho} \right)_{T}, \\
  \Gamma_{3} - 1 \equiv \left( \frac{\partial\ln T}{\partial\ln \rho} \right)_{S}.
\end{gather}
As discussed by \citealp*{Unno89} (their section 28), regions in a
stellar envelope where $\mathcal{K}$ increases outward contribute
toward a positive differential work (i.e., driving). If the outward
increase of $\mathcal{K}$ comes from a corresponding increase in the
opacity partial derivatives, $\kappa_{T}$ and $\kappa_{\rho}$, then
the driving is known as the $\kappa$ or opacity mechanism; conversely,
if it is due to an outward decrease of the $\Gamma_{3}$ adiabatic
exponent, for instance arising in an ionization zone, then it is known
as the $\gamma$ mechanism. The lower panel of \autoref{f:work-func}
plots $\mathcal{K}$ as a function of $\log T$ for the $100\,M_{\odot}$
DS model. Regions where $\mathcal{K}$ is outward-increasing (i.e., ${\rm
  d}\mathcal{K}/{\rm d}\log T < 0$) can be seen in the $\log T$
intervals $[4.70,4.55]$, $[4.30,4.25]$ and $[4.15,3.90]$. Generally,
there is good agreement between these regions and the positive peaks
in the differential work seen in the upper panel of the figure.

Further investigation reveals these regions correspond to the star's
ionization zones: the innermost is the He\,\textsc{ii} zone, the
middle is the He\,\textsc{i} zone, and the outermost the H zone. In
the He\,\textsc{ii} and H zones, the outward-increasing $\mathcal{K}$
comes from the behaviour of both $\kappa_{T}$ and $\Gamma_{3}$; hence,
the driving in these zones is via a hybrid $\kappa-\gamma$
mechanism. For the He\,\textsc{i} zone, there is almost no
contribution toward an outward-increasing $\mathcal{K}$ from either
$\kappa_{T}$ or $\kappa_{\rho}$; the driving is solely via the
$\gamma$ mechanism.

Similar results follow from examination of models at other masses. The
reason why the pulsational instability disappears above $M_{\ast}
\approx 210\,M_{\odot}$ is that the higher stellar effective
temperatures position the three $\kappa-\gamma$ driving regions so
close to the stellar photosphere that the local thermal timescale is
shorter than the pulsation period. Then, these regions remain in
thermal equilibrium over a pulsation cycle, and are unable to
contribute toward driving or damping.

All in all, we do not expect that supermassive DSs with masses far exceeding several 
hundred solar masses to pulsate by means of the $\kappa-\gamma$ and $\gamma$ mechanisms for the 
accretion rate and WIMP masses chosen in the present paper.
In the conclusion section, we briefly discuss how our results may be affected by 
other parameter choices.

\subsection{Mass Loss Induced by Pulsations}
As pulsations are excited in DSs, they provide - besides dynamical instabilities and super-Eddington winds, another potential harm that could prevent DSs from growing more massive: if the energies confined in the eigenstates of the pulsations become too high, mass outflows driven by these pulsations might occur.

We follow the analyses of \citealp*{Baraffe:2000dp} and \citealp*{Inayoshi13} to obtain an estimate for this mass loss. Making the conservative assumption that the pulsational energy is entirely transferred into the kinetic energy of the mass outflows, the mass loss rate, $\Mpuls$, can be obtained with
\beq
	\frac{\Mpuls}{2} \vesc^2 = \frac{\sigma_R}{2 \pi} W(M_\star) = 2 \sigma_I E_p \, , 
	\label{eq:massloss}
\eeq
with the real and imaginary part of the pulsation frequency, $\sigma_R$ and $\sigma_I$, respectively, the escape velocity $\vesc = \sqrt{2GM_\star/R_\star}$ and the pulsational energy of the p-mode $E_p$. The latter quantity is given by
\beq
	E_p = \frac{\sigma_R^2}{2} \int_0^{M_\star} |\delta r|^2 \, {\rm d} M_r \, ,
\eeq
where $\delta r$ denotes again the radial displacement perturbation of a fluid element at radius $r$.

As mentioned before, the linear analysis with GYRE prevents us from computing the final amplitude of unstable modes; rather we can only obtain the radial profile of the relative amplitude $\xi_r = \delta r / \delta r_{\rm surf}$. If we had observations of the pulsational velocities at the DS surface, we could normalize it via $\delta r_{\rm surf} = v/\sigma_R$. However, the lack of DS observations prohibits us from obtaining an absolute value for $E_p$. Nevertheless, the primary goal here is to determine an order of magnitude estimate of the mass loss rate that could potentially be induced by pulsations. Therefore, we use the approach of \citealp*{AppenzellerI,AppenzellerII} who showed that pulsational driven mass outflows set in, once the pulsational velocity at the surface reaches the speed of sound at the surface, $c_{\rm s}$, i.e.\! once
\beq
	\delta r_{\rm surf} = \frac{c_{\rm s}}{\sigma_R} \, 
\eeq
is fulfilled. Using this condition, the linear result for $E_p$ can be rescaled to obtain a prediction for the mass loss rate induced by pulsations. 

We calculated this upper limit for the mass loss rate according to Eq.~(\ref{eq:massloss}) for DS models with pulsations, i.e. those from $\sim 50-200 ~\MS$. In each case, the mass loss rate stays at least one order of magnitude below the accretion rate of $10^{-3} \, \MS$/yr. In our previous work RD15, we considered also higher accretion rates, in which case the mass loss would be even less significant. Therefore, we conclude that mass outflows driven by pulsations will not prevent DSs in the considered mass range from growing even more massive.

%
\section{Conclusions} \label{sec:Concl}
%
Dark stars might have been the first luminous objects to form in the Universe and provide -- through their distinctive properties compared to ordinary stars -- a laboratory to test fundamental physics. We investigated stellar physical aspects of the first dark stars to assess, whether they can be prevented from growing supermassive by potentially destabilizing astrophysical effects. 

The formation process can only take place in the early Universe ($z \sim 10-50$), where the DM density inside protostellar clouds and within the resulting stars is high enough. While ordinary stars from that era are usually too dim to be observed, DSs may grow massive and bright enough to be detected by JWST (\citealp*{Freese10, SMDSHST, Ilie12}). This crucial difference to normal stars arises from the fact that DSs are fueled by WIMP annihilation, rather than by nuclear fusion. Therefore, 
their effective temperatures are much lower which allows them to keep accreting material from their primordial surroundings at a high rate over a long period of time, without suffering feedback effects. 

A detection of very bright point sources with relatively low surface temperatures at redshift $z > 10$ with upcoming space-based and next-generation ground-based observatories could indicate the presence of DSs. We investigated the basic stellar properties of early DSs whose stability is crucial for the formation of even more supermassive DSs which might be detectable soon. 
Even though DSs are able to grow for a long time, supermassive DSs will at some point run out of DM fuel and collapse. In the process, black holes with masses of order $10^5-10^6 \MS$ might be forming, providing the seeds for the supermassive black holes seen in galaxies at high and low redshift.

In this paper, we focused on early DSs of masses $\lesssim 1000 \, \MS$ and 
found the following main results: 
\begin{itemize}
	\item The fundamental properties of DSs are not sensitive to the exact treatment of the fraction of the energy injection per WIMP annihilation, $\fq$, into the surface layers of the star. 
	\item The growth of DSs up to $\sim \!1000 \, \MS$ is not limited by dynamical instabilities.
	\item DSs of $10-1000 \MS$ are not subject to mass loss driven by super-Eddington winds. 
	\item DSs with masses $\lesssim \! 200 \, \MS$ can pulsate with periods of $\sim \! 60-600$ days in their rest frame, excited by the $\kappa-\gamma$ and $\gamma$ mechanisms in (partially) ionized layers of H and He, respectively. 
	\item The growth of DSs with pulsations is not affected by mass outflows arising from the excited radial pulsation modes. Even under conservative assumptions, the mass loss rate stays at least one order of magnitude below the accretion rate. 
\end{itemize}
%

In this work, we have not investigated whether products from WIMP annihilations in the surface layers of the collapsing baryon cloud or of the star, which cannot deposit all of their energy before leaving the object, would prevent DSs from either forming, or growing. To address this question, fully self-consistent hydrodynamical simulations of the formation and evolution of DSs, including the calculation of WIMP annihilation energy cascades, would be necessary. Such a study is beyond the scope of this work. 

However, we studied the consequences of a functional dependence of the fraction of injected energy per WIMP annihilation in the low-density layers of the stars. We applied several density thresholds $\rhot$, below which we abruptly set the luminosity infused by DM annihilation heating (Eq.~(\ref{eq:DMheatingLum})) to zero. This is motivated by the fact that below a certain baryon density the interaction cross sections between annihilation products and baryons become too small for an efficient equilibration to take place within the star. While our approach may appear a crude approximation, the results show that the exact form of $\fq$ is not a crucial factor in the evolution of DSs: DS models with different cutoff thresholds $\rhot$ differ by less than $\sim \! 1\%$ in their stellar properties, like luminosity, temperature and radius. 

Furthermore, we showed that in DSs of masses $\lesssim 200 \, \MS$ pulsation modes are excited, likely driven by the $\kappa - \gamma$ and $\gamma$ mechanisms in (partially) ionized hydrogen and helium layers. The pulsation periods of the considered models are a few hundred to a few thousand days in the observers' rest frame. Our findings are in very good agreement with the study of {\it adiabatic} pulsation periods of DSs from \citealp*{TRD15}, where it was shown that the pulsation periods of DSs decrease with increasing mass. We confirmed this trend here.


However, our non-adiabatic pulsation analysis carried out in this paper implies that, for the parameter choices made here, supermassive DSs ($\gg 10^3 \MS$) do not pulsate. However, this statement depends upon critical free parameters, namely WIMP mass and halo environment, notably accretion rate. For the parameters we chose here (100 GeV WIMP mass and $10^{-3} \MS/\rm{yr}$ accretion rate), pulsations cannot be expected in order to distinguish such DSs observationally from other sources, contrary to our initial hope expressed in \citealp*{TRD15}. By the same token, we might argue that supermassive DSs seem to be safe against the potentially disruptive power of pulsations, because standard theory expects that pulsations - if present - can turn nonlinear and violent at very high stellar masses. 

In the future, we plan to expand our studies of DS pulsations to a wider variety of parameter choices --- specifically WIMP mass and stellar accretion rate --- to 
examine whether or not supermassive DSs with different parameters than the ones chosen in this paper pulsate.
 If the WIMP mass is reduced, the DM heating is enhanced (see Eq.(\ref{eq:DMheat})), resulting in more pressure support, which in turn leads to DSs with larger radii and lower effective temperatures, see e.g. \citealp*{TRD15}. This implies that, for lower WIMP mass, a DS could be more massive at around the same ``critical'' $T_{\rm{eff}}$ for which driving is still efficient enough. For example, ``extrapolating'' our findings to the results of \citealp*{TRD15} for a WIMP mass of 10 GeV, it appears that DSs might pulsate up to many thousands of solar masses, which is still below the stellar masses observable with upcoming telescopes. Even smaller WIMP masses might lead to pulsations in very massive DSs (possibly millions of solar masses); then JWST could have the sensitivity to observe these DSs and their pulsations. This case deserves further investigation.

In addition, another effect which may alter our conclusion about DS pulsations includes the halo environment, in particular the accretion rate which in the present paper we chose to be $10^{-3} \MS/\rm{yr}$. Implications for pulsations and whether they may be excited, or not, due to the accretion rate have been found for supermassive ``protostars'' in \citealp*{Inayoshi13}, who found that stars of mass $M_{\star} \gtrsim 600 \MS$ are pulsationally unstable due to the $\kappa$ mechanism, if they grow at very high accretion rates $\gtrsim 1.0 ~\MS/\rm{yr}$, and that their pulsations are excited in layers of singly-ionized helium. Similarly, rapidly accreting supermassive protostars with masses up to $10^5 \MS$ have been found to exhibit pulsational instabilities in \citealp*{Hosokawa13}. In fact, we know that, for a given WIMP mass, the sharp increase in $T_{\rm{eff}}$ mentioned above happens later - i.e. at a higher stellar mass, if the accretion rate is increased (compare to Figs. 2,8,9 in \citealp*{TRD15}). The difference which results is more pronounced for high WIMP mass, but seems of less relevance for smaller WIMP mass. Hence, there is a multitude of parameters (WIMP mass, halo environment) which could conspire to produce pulsational instabilities for supermassive DSs, after all. However, in order to settle this question, we would need to re-do our analysis for different values of WIMP mass and accretion rate, a project which we defer to future work.

On the other hand, we considered the $\kappa$ and $\gamma$ mechanisms as the only mechanisms to drive pulsations. The $\epsilon$-mechanism, where modulations in the core temperature lead to perturbations in the nuclear fusion rates, is not present in DSs solely powered by the heating of DM annihilations. However, perturbations in the energy injection rate in DSs could arise from variations in the DM density, affecting the luminosity, Eq.~(\ref{eq:DMheatingLum}), through a modulation of $\Qdm$. This ``$\chi$-mechanism'' driven by DM may possibly also excite pulsations in DSs. We leave an analysis of this effect and a study of supermassive DSs to future work.

\section*{Data availability statement}
All data are incorporated into the article and its online supplementary material.

\section*{Acknowledgements}

We are deeply indebted to Bill Paxton for extensive support with MESA. We thank Janina Renk and Sebastian Baum for their helpful input. Furthermore, we thank Michael Montgomery for discussions and suggestions to improve our work. T.R.-D. acknowledges the financial support of the Austrian Science Fund FWF through an Elise Richter fellowship, grant nr. V656-N28. T.R.-D. acknowledges the financial support of the FWF through a Lise Meitner fellowship, grant nr. M2008-N36, during the early stages of this work. 
K.F. is Jeff \& Gail Kodosky Endowed Chair in Physics at the University of Texas at Austin, and is grateful for support.
K.F. acknowledges support from the U.S. Department of Energy, grant DE-SC007859 and the Leinweber Center for Theoretical Physics at the University of Michigan. K.F. and L.V. acknowledge support by the Swedish Research Council (Contract No. 638-2013-8993). 
L.V. thanks the Leinweber Center for Theoretical Physics at the University of Michigan, where part of this work was conducted, for hospitality. L.V. acknowledges support from the NWO Physics Vrij Programme ''The Hidden Universe of Weakly Interacting Particles'', nr.680.92.18.03, which is partly financed by the Dutch Research Council NWO.
R.H.D.T. acknowledges support from the U.S. National Science Foundation under awards OAC-1663696 and AST-1716436.\\
The stellar evolution calculations presented in this paper were made using MESA 12778 (\citealp*{NewMESA}). The non-adiabatic pulsation calculations were made using GYRE (\citealp*{Gyre,GT20}).







\appendix

\section{Equations for adiabatic contraction}\label{sec:Appendix}

We present the equations used to calculate the adiabatically contracted DM density profile after baryonic cooling and infall via the Blumenthal method (\citealp*{Blumenthal}). This part is crucial, since the energy source of DSs directly depends on this quantity (see Eqs.~(\ref{eq:DMheat}-\ref{eq:DMheatingLum})). The implementation of the DS heating routine into MESA requires not only the formula for the heating rate per unit mass, but also its partial derivatives with respect to the stellar density $\rhos$, radius $r$ and temperature $T$. Notice that the temperature derivative is irrelevant and can be set to zero in DSs, because the heating rate only depends upon the DM and baryonic densities and is independent of the temperature, see Eqs.~(\ref{eq:DMheat}-\ref{eq:DMheatingLum}).

In the previous implementation of RD15, the derivatives with respect to radius and stellar density were implemented as numerical difference equations. In this work, in order to improve numerical stability, we replaced these numerical derivatives with analytical expressions. We first review the calculation of the adiabatic contraction (AC) with the Blumenthal method and then give the expressions for the heating rate and its derivatives.

Let us assume that the mass profile of contracted baryons is given by $M_b(r)$, where $r < r_i$ is the final radius of the particles initially located at $r_i$ before contraction. According to the assumption of adiabatic infall in a spherical potential, the baryon profile satisfies (see Eq.~(1) in \citealp*{Blumenthal})
\begin{equation} \label{adiab_contr}
r_i\,M_i(r_i) = \left[(1-f)\,M_i(r_i) + M_b(r)\right] r \,,
\end{equation}
where $f = M_b(\rv)/\Mv$ is the fraction of baryons contributing to the total mass at the virial radius, $\rv$.
The DM mass at radius $r$ is then $M_\chi(r) = (1-f)\,M_i(r_i)$. To solve Eq.~(\ref{adiab_contr}) we follow \citealp*{Gnedin} and assume 
that the contracted baryon profile scales as $M_b(r) \propto r^2$ for $r \ll \rv$, as in the exponential disk model describing spiral galaxies (\citealp*{Freeman:Disk}).
Then, $M_b(r) = M_b(r_i)/y^2$, where $y = r_i/r$. For the DM, we take an initial NFW profile, see Eq.~(\ref{eq:iniNFWprof}).

Dividing Eq.~(\ref{adiab_contr}) by $M_i(r_i)\,r$ gives
\begin{equation} \label{adiab_contr1}
	y = (1-f) + \frac{M_b(r_i)}{y^2 M_i(r_i)} \equiv (1-f) + \frac{B}{y^2} \, ,
\end{equation}
where the parameter $B$ can be considered as a constant here since in both profiles, NFW and exponential disk, the masses scale with $r^2$ for $r \ll \rv$. The only real solution to the cubic Eq.~(\ref{adiab_contr1}) reads
\begin{eqnarray} 
r_i &=& \frac{1}{3}\,\left[c_1 + \frac{c_3}{2^{1/3}} + \frac{2^{1/3} \,c_1^2}{c_3}\right] \, {\rm with} \label{eq:r_i}\\
c_1 &=& r(1-f),\nonumber\\
c_2 &=& r^3\,B= \frac{M_b(r)\,r}{2\pi\,\rho_0\,r_s}, \nonumber\\
c_3 &=& \left(27c_2+2c_1^3 + 3\sqrt{3 e_3}\right)^{1/3},\nonumber\\
e_3 &=& 4 c_1^3c_2 + 27 c_2^2 \,  \nonumber.
\end{eqnarray}
%

With this expression for $r_i$, the DM mass enclosed within $r_i$ can be approximated by $M_i(r) \approx \Mv\,\frac{r^2}{2\theta\,r_s^2}$ for $r \ll r_s$. The DM mass profile is then given by
\begin{equation} \label{tot_DMmass}
M_\chi(r) = (1-f)\,M_i(r_i) = (1-f)\,(2\pi\,\rho_0\,r_s)\,r_i^2 \, ,
\end{equation}
where
\begin{equation}
\rho_0 = \frac{\rhv\,c^3}{3\theta} = \frac{\Delta_c\,\rho_{\rm{crit}}(z)\,c^3}{3\theta} \, 
\end{equation}
denotes the characteristic density (compare to Eq.(\ref{eq:iniNFWprof})).
Now, the DM density can be expressed as
\begin{equation} \label{tot_DMrho}
\rhodm = \frac{ \der_r M_\chi}{4\pi \,r^2} = \frac{(1-f)\,\rho_0\,r_s}{2r^2}\, \der_r (r_i^2) = \frac{(1-f)\,\rho_0\,r_s}{r^2}\,r_i\, \der_r r_i \, ,
\end{equation}
where $\der_Y X \equiv \der X / \der Y$ and we omitted the radius $r$ as an argument of $\rhodm$. Notice that $\rhodm$ depends explicitly on both, the stellar radius and density, $r$ and $\rhos$. The dependence on $\rhos$ comes from taking the derivative
\begin{eqnarray} \label{eq:dc2dr}
\der_r c_2 &=& \frac{M_b(r)+4\pi\,r^3\,\rhos}{2\pi\,\rho_0\,r_s},
\end{eqnarray}
so that the derivative of Eq.~(\ref{eq:r_i}) reads
\begin{equation} \label{eq:dridr}
\der_r r_i = \frac{1}{3}\left[(1 - f)\,\left(1 + \frac{2^{4/3}\,c_1}{c_3}\right) + \left(\frac{1}{2^{1/3}} - \frac{2^{1/3} c_1^2}{c_3^2}\right) \der_r c_3 \right] \, ,
\end{equation}
with 
\begin{eqnarray} \label{eq:dc3dr}
\der_r c_3 = \frac{4 c_1^2 (1-f)+ 18 \, \der_r c_2 \sqrt{3/e_3} \, \der_r e_3}{2c_3^2}\, , \nonumber\\
\der_r e_3 = 12 c_1^2 c_2 (1-f) + (4 c_1^3 + 54 c_2) \,  \der_r c_2 \, .
\end{eqnarray}

Now, with the expressions for $r_i$ and $\der_r r_i$, $\rhodm$ and therefore the heating rate \textit{per unit mass},
\begin{equation} \label{Qm_unit_volume}
Q_m = \frac{\langle\sigma\,v\rangle\,\rho_\chi^2}{m_\chi \, \rhos} 
\end{equation}
are fully specified. The partial derivatives of the latter, needed to consistently include the calculation of the extra energy source in MESA, read
\begin{displaymath}  
\partial_{\ln r} Q_m = 2 Q_m \frac{\rhos }{ \rhodm} \, \partial_r \rhodm = 2 Q_m \, \frac{\rhos}{\rhodm} \frac{(1-f)  \rho_0 r_s}{ r^2}
\times
\end{displaymath}
\begin{equation}  \label{eq:parQ}
\times \left[-2 \, \frac{r_i}{r} \der_r r_i + \partial_r r_i \der_r r_i + r_i \parrdr{r_i} \right] \, ,
\end{equation}
\begin{displaymath}
\partial_{\ln \rhos} Q_m = Q_m \left[ \frac{2\, r}{\rhodm} \, \partial_{\rhos} \rhodm -1 \right]  
\end{displaymath}
\begin{equation}   \label{eq:parQ1}
=  Q_m \left[ \frac{2\, r}{\rhodm} \, \frac{(1-f) \rho_0 r_0}{r^2} \left( \partial_{\rhos} r_i  \,\der_r r_i + r_i \, \parddr{r_i} \right) -1 \right] \, ,
\end{equation}
where $\partial_Y X \equiv \partial X/ \partial Y$. Calculating these expressions requires the partial derivatives of $r_i$ and $\der_r r_i$ with respect to
$r$ and $\rhos$. The expression for $\partial_r r_i$ can be obtained by simply replacing $\der_r \rightarrow \partial_r$ in Eqs.~(\ref{eq:dridr}-\ref{eq:dc3dr})
for $\der_r r_i$, while the partial derivative of $c_2$ reads 
\beq \label{eq:parc2dr}
\partial_r c_2 = \frac{M_b(r)}{2\pi\,\rho_0\,r_s} = c_2 / r \, .
\eeq
The partial derivative of $\der_r r_i$ with respect to $r$ amounts to

\begin{displaymath}  
\parrdr{r_i} = \frac{1}{3c_3^2} 
\bigg[ 2^{4/3} (1-f)^2 \left(c_3 - \frac{c_1 \, \partial_r c_3}{1-f} \right) +
\end{displaymath}
\begin{equation} \label{eq:parrdr_ri}
            + ~ 2^{4/3} c_1 \, \der_r c_3 \frac{ c_1 \, \partial_r c_3 - c_3(1-f)}{c_3} 
            + \frac{\partial_r \, (\der_r c_3)}{ 2^{1/3}} \left(c_3^2 - 2^{2/3} c_1^2 \right) \bigg] \, ,
\end{equation}
with
\begin{displaymath}
\parrdr{c_2} =  6 \, \frac{ \rhos r^2}{\rho_0 r_s} \, ,
\end{displaymath}
\begin{displaymath}
\parrdr{c_3} = \frac{1}{2 c_3^2} \left[8 c_1 (1-f)^2 + 18 \, \parrdr{c_2} + \right. 
\end{displaymath}
\begin{displaymath}
\left. + ~ \sqrt{3/d_3} \left( \parrdr{e_3} - \frac{\der_r e_r \, \partial_r e_3}{2 e_3}\right) \right] - 2 \, \frac{\der_r c_3 \, \partial_r c_3}{c_3}
 \, ,
 \end{displaymath}
 \begin{displaymath}
 \parrdr{e_3} = 12 c_1 (1-f) \big[ 2 (1-f) c_2 + c_1 \partial_r c_2 \big] + 
 \end{displaymath}
 \begin{displaymath}
 + \left[12 (1-f) c_1^2 + 54 \partial_r c_2 \right] \, \der_r c_2 + \big(4 c_1^3 + 54 c_2 \big) \, \parrdr{c_2} \, .
\end{displaymath}
The partial derivative of $r_i$ with respect to $\rhos$ is zero, because $r_i$ has no explicit density dependence, see Eq.~(\ref{eq:r_i}). However, $\parddr{r_i}$ does not vanish, because the density dependence of $\der_r r_i$ enters through the total derivative of $c_2$, see Eq.(\ref{eq:dc2dr}). The expression for $\parddr{r_i}$ can be obtained by replacing $\partial_r \rightarrow \partial_{\rhos}$ in Eq.~(\ref{eq:parrdr_ri}) and by using
\begin{eqnarray}
\parddr{c_2} &=& 2 \frac{r^3}{\rho_0 r_s}  \, , \nonumber\\
\parddr{c_3} &=&  \frac{18 \, \parddr{c_2} + \sqrt{3/e_3} \, \parddr{e_3}}{2 c_3^2}     \, , \nonumber\\
\parddr{e_3} &=& (4 c_1 + 54 c_2) \, \parddr{c_2} \, .
\end{eqnarray}

Using these equations, the partial derivatives of the heating rate, Eq.~(\ref{eq:parQ}) and Eq.~(\ref{eq:parQ1}), are fully specified.


\bsp	
\label{lastpage}
\end{document}